\documentclass[aps,prb,footinbib,showpacs,twocolumn,showkeys,amssymb,amsmath,superscriptaddress]{revtex4-1}
\pdfoutput=1
\usepackage{graphicx,graphics}
\usepackage[export]{adjustbox}
\usepackage{verbatim}
\usepackage[font=small]{caption}
\usepackage{braket}
\usepackage{float}
\usepackage{breqn}
\usepackage{subfig}
\usepackage{color}
\usepackage{hyperref}
\usepackage{enumerate}
\makeatletter
\let\cat@comma@active\@empty
\makeatother

\begin{document}
	\title{Topological quantum phase transitions and criticality in a longer-range Kitaev chain}
	
	\author{Y. R. Kartik}
	\thanks{yrkartik@gmail.com}
	\author{Ranjith R. Kumar}
			\thanks{ranjith.btd6@gmail.com}
	\author{S. Rahul}
			\thanks{rahulastronomer02@gmail.com}
	\affiliation{Theoretical Sciences Division, Poornaprajna Institute of Scientific Research, Bidalur,	Bengaluru, 562164, India.}
	\affiliation{Graduate Studies, Manipal Academy of 
	Higher Education, Madhava Nagar, Manipal-576104, India.}
	\author{Nilanjan Roy}
				\thanks{nilanjanroy.physics@gmail.com}
	\affiliation{Department of Physics, Indian Institute
	 of Science Education and Research, Bhopal, Madhya 
	 Pradesh 462066, India}
     \affiliation{Department of Physics, Indian Institute of Science, Bangalore 560012, India}
	\author{Sujit Sarkar}
		 \thanks{sujit.tifr@gmail.com}
		\affiliation{Theoretical Sciences Division, Poornaprajna Institute of Scientific Research, Bidalur,	Bengaluru, 562164, India.}

	\date{\today} 
	
\begin{abstract}
In an attempt to theoretically investigate the quantum phase transition and criticality in topological models, we study Kitaev chain with longer-range couplings (finite number of neighbors) as well as truly long-range couplings (infinite number of neighbors). We carry out an extensive topological characterization of the momentum space to explore the possibility of obtaining higher order winding numbers and analyze the nature of their stability in the model. The occurrences of phase transitions from even-to-even and odd-to-odd winding numbers are observed with decreasing longer-rangeness in the system. We derive topological quantum critical lines and study them to understand the behavior  of criticality. A suppression of higher order winding numbers is observed with decreasing longer-rangeness in the model. We show that the mechanism behind such phenomena is due to the superposition and vanishing of the topological quantum critical lines associated with the higher winding number. Through the study of Berry connection we show the possible different behaviors of critical lines when they undergo superposition along with the corresponding critical exponents. We analyze the behavior of the long-range models through the momentum space characterization. We also provide exact solution for the problem and discuss the experimental aspects of the work.
\end{abstract}
	
\maketitle
	
\section{Introduction}
Topological states of matter are considered as novel 
phases of matter in modern physics. The concept 
started as a theoretical prediction and expanded
 towards experimental realizations~\cite{hasan2010colloquium,moore2010birth}. 
 For almost a century, Landau theory of 
 spontaneous symmetry breaking explained
  almost all the phases of matter except 
  topological phases~\cite{landau1965collected,
  miransky1993dynamical,vojta2003quantum}. 
  Landau theory relies on the existence of the local
   order parameter, which is absent in topological
   state of matter~\cite{wen2016introduction}. 
   This created a need for some alternate way 
   to establish topological characterization. 
  Topological invariant is a promising quantity, 
  which explains gapped topological phases in a 
  very accurate way, but it fails when comes to 
  topological quantum phase transition (TQPT)~ \cite{bohm2013geometric}.
   
TQPT is basically quantum phase transition at 
quantum critical points (QCP), since they occur 
at zero temperature~\cite{ortmann2015topological,stanescu2016introduction,bernevig2013topological}.
 At QCP, instead of vanishing local order parameter,
  topological system has a special kind of diverging
   topological correlation factor in its electronic 
   band structure~\cite{continentino2020finite}. 
   This topological correlation factor is directly
    associated with topological invariant. Hence 
    topological invariants are well quantized at 
    the gapped phases and ill-defined at QCP. So 
    far for a 1D system, winding number (WN) is the most accepted form
     of topological invariant which is the integration
      of Berry connection (vector potential) over the 
      Brillouin zone. Hence considering this form of the
 correlation factor, it is possible to extract the 
 information around the QCP \cite{rufo2019multicritical}. This insight helped 
 scientific community to think about topological state of matter from the 
 perspective of criticality. Even earlier  there 
 were many attempts like renormalization group~
 \cite{sachdev2007quantum,kumar2020quantum,sarkar2020study}, curvature function 
 renormalization group~\cite{chen2019topological}
  and other scaling approaches to explain TQPT~\cite{abasto2008fidelity,amin2019information}.
 But now it is evident that there is a possibility
 to explain criticality through correlation function, curvature function~\cite{chen2019topological}, 
 critical exponents and universality class of 
 TQPT~\cite{continentino2020finite,rufo2019multicritical}.
 
Topological states of matter are the area of curiosity because of the emergence 
of exotic quasi-particles unlike fermions and bosons~\cite{kitaev2001unpaired}. The area became more prominent 
with its real life applications~\cite{sarma2015majorana}. 
It is possible to generate higher order localized edge 
modes through periodic driving and longer-range 
couplings, 
where the previous method yields dynamical localized modes 
and the later yields static modes~\cite{cats2018staircase,tong2013generating}. 
Higher order localized modes have their own 
interests in topological state of matter. On the other hand long-range topological models 
are the more generalized version of novel phases 
of matter \cite{vodola2014kitaev}. This includes realization of new phases like edge insulating 
topological 
phases~\cite{viyuela2018chiral,lepori2018edge} with fractional topological 
invariants \cite{alecce2017extended} and
quasi-particles 
like Majorana zero modes (MZM), massive majorana modes~\cite{viyuela2016topological}.
In this work, we carry out a theoretical study of a topological longer-range as well as truly long-range model. Our motivation is two folded: topological characterization of the momentum 
space and study of the quantum criticality for the long-range models.

We theoretically study the topological quantum phase transition and criticality in Kitaev chain with `longer-range' couplings which are finite-ranged but more than nearest-neighbors. Here we stick to the terminology used in the existing literature~\cite{alecce2017extended,niu2012majorana}. We derive topological quantum critical lines and find a suppression of higher order winding numbers with 
decreasing longer-rangeness in the model. As a reason behind this we argue 
for the superposition of two critical lines with different 
winding numbers followed by vanishing of the critical line 
with higher winding number. 
 We analyze the possibilities of obtaining higher order winding numbers and study their stability in the model. Our analysis shows the phase transition from even-to-even and odd-to-odd winding numbers with decreasing longer-rangeness in the system.
We also provide a few exact solutions for winding number.

This paper is organized as follows. In Section
~\ref{Ham} we explain our model Hamiltonian and aim of the study. In Section~\ref{winding} we carry out the topological characterization of momentum space by calculating winding number. We also obtain the phase diagram with a detailed study of the critical lines.  
In Section~\ref{char}, We elaborate on the topological quantum criticality of our longer-range system and extend it with respect to stability of higher order winding numbers. We perform momentum space characterization to explain the behavior of long-range models. In Section~\ref{para} we study the parameter space in order to understand critical phases with a few relevant exact solutions.
In Section~\ref{gen} we provide the outlook and experimental aspects of the work. Then we conclude in Section~\ref{conclusion}.

\section{The model and aim of the study}\label{Ham}
We consider 1D Kitaev model with $r$ neighboring interactions 
(both hopping and pairing) ~\cite{alecce2017extended}. This kind of model for infinitely long-range model was studied in Ref ~\cite{vodola2014kitaev} and for longer-range Kitaev chain was studied in Ref ~\cite{alecce2017extended}. This model helps to understand the emergence and behavior of Majorana modes and topological invariant in the longer-range as well as in the truly long-range systems. We define our model Hamiltonian as,
\begin{equation}
H=-\sum_{j=1}^{L}\mu(c_j^{\dagger}c_j-1/2)-\sum_{j=1}^{L-l}
\sum_{l=1}^{r}(J_lc_j^{\dagger}c_{j+l}+\Delta_lc_j^{\dagger}
c_{j+l}^{\dagger}+H.c.),\label{eq}
\end{equation}
where $\mu$ is the chemical potential, $L$ is the number of 
lattice sites, $J_l$ and $\Delta_l$ are the strengths of hopping and 
pairing terms respectively with long-range interactions of the form
$$J_l=\frac{J_0}{d_l^{\alpha}}, \Delta_l=\frac{\Delta_0}
{d_l^{\beta}}.$$ 
These hopping and pairing terms couple the lattice site $j$ 
with $j+l$. For a system with open boundary condition, the distance 
$d_l=l$. For closed boundary $d_l=min(l,L-l)$.
$\alpha$ and $\beta$ are the non-negative parameters which represent the power-law decay of hopping and pairing terms respectively. When  $\alpha\rightarrow\infty$, 
system behaves as a Kitaev chain with only long-range pairing and when 
$\beta\rightarrow\infty$, system behaves as a Kitaev chain with only long-
range hopping. When both 
$\alpha,\beta\rightarrow\infty$ 
system behaves as original Kitaev chain~\cite{kitaev2001unpaired,vodola2014kitaev,alecce2017extended}.

After a Fourier transformation, one can write 
the model in the momentum space as
	\begin{dmath}
		H  =   \sum_{k> 0} \left(-\mu-2J_0\sum_{l=1}^{r}\frac{\cos[kl]}
		{l^{\alpha}}\right)
	({\psi_k}^{\dagger} {\psi_k} + {\psi_{-k}}^{\dagger} 
	{\psi_{-k}})
	+  2i \Delta_0  \sum_{k > 0} \sum_{l=1}^{r}\left(\frac{\sin[kl]}{l^{\beta}}\right)
	({\psi_k}^{\dagger} {\psi_{-k}}^{\dagger} +
	{\psi_{k}} {\psi_{-k}}),\end{dmath}
	 where ${\psi^{\dagger}}(k)$ $(\psi (k))$ is the 
	creation (annihilation) operator of the spinless fermion 
	of momentum $k$. 		
	We can write the BdG Hamiltonian  as 
	\begin{equation}
	H_{BdG}(k)=\left(\begin{matrix}
	\chi_z(k)&& i\chi_y(k)\\
	-i\chi_y(k)&& -\chi_z(k)\\
	\end{matrix}
	\right).\end{equation}\label{matrix}	
	We can express the Hamiltonian by Anderson 
	pseudo-spin approach~\cite{anderson1958coherent,sarkar2018quantization}. 
	One can write the BdG Hamiltonian in the pseudo-spin basis as 
	\begin{dmath}
	H_{BdG}(k)=\chi_x(k)\vec{\tau_1}+\chi_y(k)\vec{\tau_2}
	+\chi_z(k) \vec{\tau_3}, 
	\end{dmath}\label{pseudo}
	where ${\tau_i}=(\tau_1,\tau_2,\tau_3)$ 
	 are the Pauli matrices in particle-hole space and the coefficients are
	 \begin{eqnarray}
	 \chi_x(k)&=& 0,\nonumber\\ \chi_y(k)&=& 2\Delta_0 \sum_{l=1}^{r}\frac{\sin 
	 [kl]}{l^{\beta}},\nonumber\\
	 \chi_z(k)&=&(-\mu-2J_0\sum_{l=1}^{r}\frac{\cos[kl]}{l^{\alpha}}).
	 \label{pseudospin}
	 \end{eqnarray} 
It is to be noted that for $r\rightarrow\infty$ the series involving $\frac{\cos(kl)}{l^{\alpha}}$ and 
$\frac{\sin(kl)}{l^{\beta}}$ terms give rise to polylogarithmic functions \cite{alecce2017extended,vodola2014kitaev}. The 
quasi-particle excitation energy is given by
\begin{equation}
E_k=\pm\sqrt{(\chi_z(k))^2+(\chi_y(k))^2}.
\label{endisp}
\end{equation}
In the current work our interest is to analyze the Kitaev model with 
finite number of interacting neighbors. We consider 
the Hamiltonian with longer-range hopping and pairing 
up to finite $r$ neighbors with $\alpha=\beta$ and 
$J_0=\Delta_0=\lambda$, so that within this regime the 
system resembles isotropic Kitaev chain with $r$ 
neighboring interactions. Hence as one varies the value 
of $r$, it is possible to generate Kitaev chain 
whose neighboring terms have a power law 
decay in the associated couplings.

Here we explain the aim of our work. 
(i) For a longer-range Kitaev model with $r$ 
nearest neighbors, there exists $r$ topological phases~\cite{cats2018staircase,alecce2017extended} 
and  one can recover 
original Kitaev chain when 
$\alpha,\beta\rightarrow\infty$,~\cite{kitaev2001unpaired,vodola2014kitaev}. 
It is a very effective method to choose more number of 
neighbors 
to achieve higher order WNs. Here our interest is to understand and analyze the possibility of obtaining all the $r$ topological 
phases in an isotropic Kitaev chain.
(ii) For a transition between topological phases of higher WN and 
lower WN, there may exist a staircase of  TQPTs~\cite{cats2018staircase}. Here we attempt to carry out an analysis to extract the order of such TQPTs and study the
stability of these high-WN topological phases.
(iii) For a longer-range model with $r$ interacting neighbors, there exist $r$ topological 
phases and critical 
lines which distinguish the topological phases \cite{niu2012majorana}. 
Hence, when a longer-range model is reduced to its original short-range 
version, there may be a change in the behavior 
of its corresponding critical lines also.
To explain this, we derive all possible critical lines 
and study their behavior from the perspective of quasi-particle 
energy spectrum, curvature function and ground-state energy. (iv) In a truly long-range model, there may exist some emergent quasi-particles like massive edge modes, based on the selection of parameter space. However, we try to analyze this phenomenon from the perspective of momentum space characterization.

\section{Topological characterization in momentum Space and Topological Quantum Criticality}\label{winding}
Winding number is the most accepted form of topological 
invariant~\cite{wilczek1989geometric,berry1985classical}. 
In this section, we 
derive WN for different choices of parameters  
to understand the possible topological index of the 
system. We explore the superposition and vanishing of TQCLs through the study of Berry connection and ground-state energy. We study 
and analyze 
the ground state energy to explain the stability of
higher order WNs~\cite{kempkes2016universalities,cats2018staircase,chen2008intrinsic,mahyaeh2018zero}. We study the parameter space through pseudo-spin vectors and also derive quite a few exact solutions for the WN. 
\subsection*{Winding number}
For a system in 1D, WN is defined as,
\begin{equation}
W=\frac{1}{2\pi}\oint\frac{\partial\theta_k}{\partial k}dk=\frac{1}{2\pi}\oint
\frac{\chi_z\partial_k\chi_y-\chi_y\partial_k\chi_z}{\chi_z^2+\chi_y^2}dk,
\label{w1}
\end{equation} 
 where 
$\theta_k=\tan^{-1}\left(\frac{\chi_y}{\chi_z}\right)$. 
This relation holds good even for longer-range models with 
$r$ nearest interacting neighbors, although this definition of $W$ is ill-defined at TQPT. In all possible gapped phases, 
topological index secures a quantized value (integers 
like $W=0,1,2,3...,r$) and this depends on the number of 
interacting neighbors. This is because, the WN
 is always associated with the modulo of $2\pi$. 
The phase $W=0$ represents non-topological phase. One can 
achieve higher order WNs by increasing the 
number of interacting neighbors. The transition from one 
topological phase to other occurs through topological 
quantum critical lines (TQCL)~\cite{sarkar2018quantization}.
 These TQCL are the gap closing points in the 
 quasi-energy spectrum. It is important to note that, 
 for all TQCLs there are the gap closings but all the gap closings 
 need not be TQCLs~\cite{chen2019topological}. The gap 
 closing results in the QPT and if 
 this QPT differentiates two distinct topological phases, 
 then this gap closing points are known as topological 
 quantum phase transition (TQPT) points.
 
WN always corresponds to the number of localized edge 
modes of the topological gapped phases. Recently there 
are some works which show the localized edge modes even 
at the criticality~\cite{jones2019asymptotic,verresen2019gapless,thorngren2020intrinsically,verresen2020topology,verresen2018topology,kestner2011prediction,cheng2011majorana,fidkowski2012majorana,sau2011number,kraus2013majorana,scaffidi2017gapless,jiang2018symmetry,kumar2020multi,rahul2019anomalous}. Hence it is clear and meaningful to 
find the WN around criticality and physically it is 
possible to find the corresponding edge modes. There are 
some cases, where one can get the fractional WNs at critical points. 
Even though, there are no proper experimental evidences 
for fractional edge modes, it is possible to define the 
fractional WN around criticality. The definition of WN for TQPT
can be modified by excluding an infinitesimal 
neighborhood of the gapless/critical points. Then the 
modified expression for WN is given by~\cite{verresen2020topology}
\begin{equation}
W=\frac{1}{2\pi}\lim_{\delta\rightarrow 0}\int_{\forall i:|k-k_i|>\delta}\frac{\partial\theta_k}{\partial k}dk,\label{fw}
\end{equation}
where $\{k_i\}$ is the set of critical/gapless points in the momentum space. Thus, we can define the 
WN at and around the critical point.

Here we consider a limited number of interacting 
neighbors $r=2,3,4$ and obtain the possible topological phases with integer WNs. The TQPTs among these topological phases for $r=2,3,4$ are shown in following cases respectively.
 We also calculate all possible critical lines 
to understand the phase diagram of the system and its dependence on the decay parameter $\alpha$.
In quantum systems, the transition occurs from one phase 
to another through the QCP, which are the gapless points 
in the excitation energy spectrum. The
quasi-particle excitation energy spectrum of our model is given by Eq.~\ref{endisp}.
\subsubsection*{Case 1: When $r=2$}   
\begin{widetext}

 \begin{figure}[H]
  	\centering
  	\includegraphics[width=14cm,height=12cm]{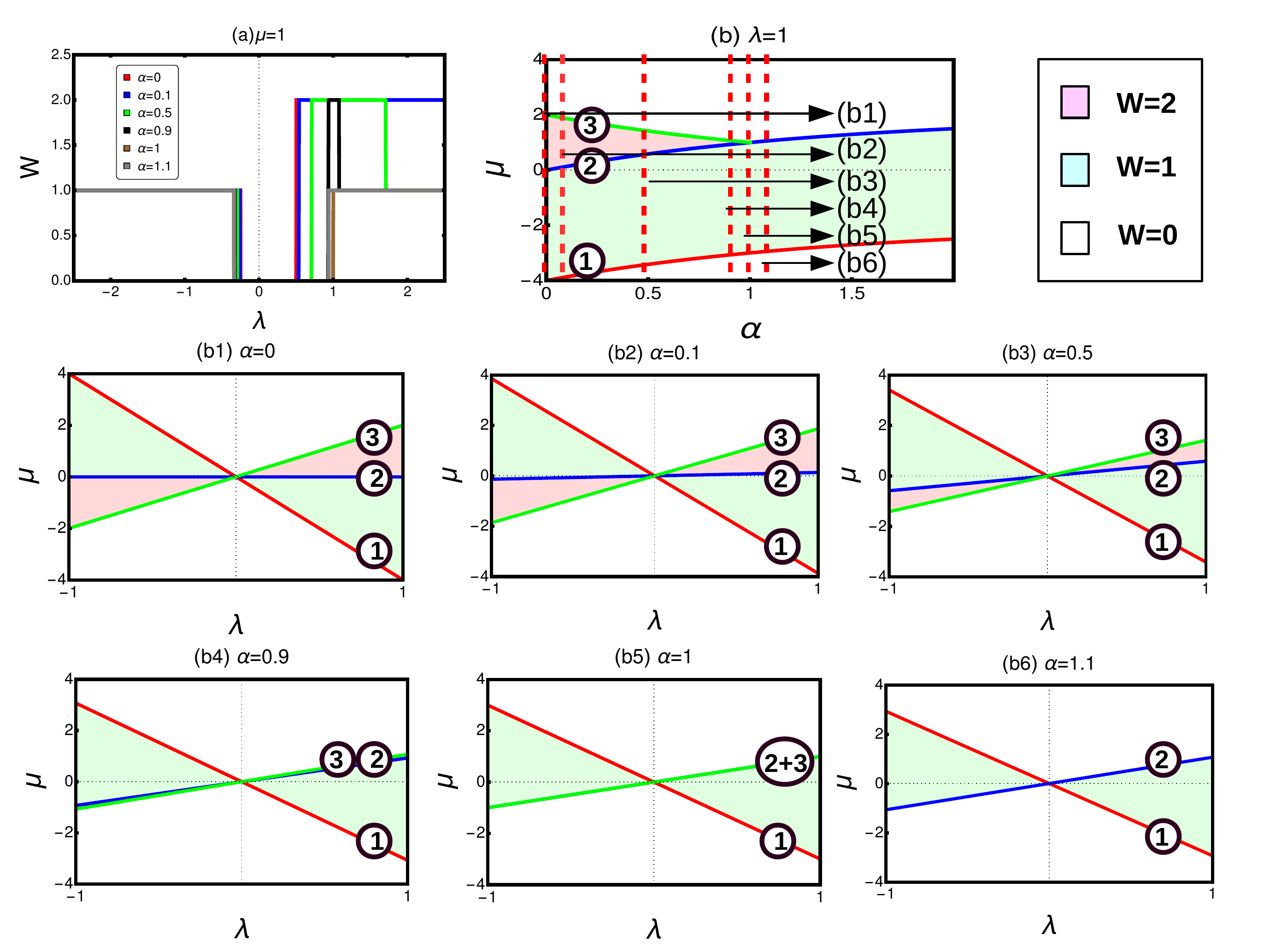}
  	\caption{ Topological phase diagram of longer-range Kitaev chain with $r=2$ neighbors. a) Winding number study with respect to $\lambda$ for different values of $\alpha$. (b) Behavior of topological phases in $\mu-\alpha$ parameter space with fixed $\lambda$. (b1-b6) Corresponding topological phase diagrams of model in $\mu-\lambda$ parameter space with increasing values of $\alpha$. This way of representation helps to understand the behavior of TQCLs.  Red, blue and green lines represent the 1st, 2nd and 3rd TQCLs respectively. With the increasing value of $\alpha$, the 2nd TQCL starts moving in anticlockwise manner to superpose with 3rd TQCL. This causes the suppression of $W=2$ phase. The 3rd TQCL vanishes for $\alpha>1$ and the model reduces to short-range Kitaev chain.}
  	\label{r2pd}
  \end{figure}
 
\end{widetext}

Fig.~\ref{r2pd} shows the phase diagram for $r=2$. Here we can 
  see the interaction is up to the second nearest neighbor in 
  the chain and Hamiltonian is given by Eq. \ref{eq}, with $r=2,\lambda=\Delta=J$ and $\alpha=\beta$. Here, the gap closings occur at three 
different values of $k$. (Appendix~\ref{appC}) which corresponds to three different 
TQCLs.

For this $r=2$, WN is given by
\begin{equation}
W=\frac{1}{2\pi}\int_{-\pi}^{\pi}\frac{\left( \frac{4 
\lambda ^2 Q A}{P^2}+\frac{2 \lambda  B}{P}\right) 
}{\left( \frac{4 \lambda ^2 Q^2}{P^2}+1\right) }dk,
 \label{wn1}
\end{equation}
 where 
 \begin{eqnarray}
A&=&\left(-2^{1-\alpha } \sin (2 k)-\sin (k)\right),\nonumber\\
B&=&\left(2^{1-\alpha } \cos (2 k)+\cos (k)\right),\nonumber\\
P&=&\left(-2 \lambda  \left(2^{-\alpha } 
\cos (2 k)+\cos (k)\right)-\mu \right), \nonumber\\
Q&=&\left(2^{-\alpha 
} \sin (2 k)+\sin (k)\right)\nonumber
 \end{eqnarray}

 Fig.~\ref{r2pd}(a) shows the WN for the case 
 $r=2$. For $\alpha=0,0.1$, one can observe transitions 
 among $W:0 \leftrightarrow 1$ and $W:0\leftrightarrow2$. For $\alpha=0.5$, one can observe the transitions 
   among $W:0\leftrightarrow1$, $W:2\leftrightarrow1$ and $W:0\leftrightarrow2$. We note that the plateau of 
 $W=2$ region for $\alpha=0.5$ is reduced as compared to that for $\alpha=0$. This indicates the decrease in the stability of the higher order WN as $\alpha$ increases. For $\alpha=0.9$, the $W=2$ topological phase has very short plateau and for $\alpha=1,1.1,$ it is clearly seen that $W=2$ region is absent and we observe only $W=1$ and $W=0$ phases similar to the original Kitaev chain.
 
 Here we have three TQCLs i.e., The red (1st TQCL for $k=0$), blue (2nd TQCL for $k=\pi$) and green lines 
   (3rd TQCL for $k=\cos^{-1}(-2^{\alpha-1})$) (see Appendix~\ref{appC}).
   Throughout the case 1st TQCL (red) separates 
  $W=1$ and $W=0$ and it is unaltered with variation of 
  $\alpha$. When $\alpha=0$, the 2nd critical line (blue) lies on 
  the $\lambda$ axis (Fig.~\ref{r2pd}(b1)). As one 
  gradually increases $\alpha$ the 2nd critical line 
  starts moving in anticlockwise direction (Fig.~\ref{r2pd}(b2-b4)) and superposes with 3rd critical line (green) 
  (Fig.~\ref{r2pd}(b5)). This results in the vanishing of 
  $W=2$ topological phase which in turn leads to vanishing of the 3rd TQCL. We observe that 
  the 3rd TQCL vanishes for $\alpha>1$ (Fig.~\ref{r2pd}(b6)), which is consistent with the fact that the point $k=\cos^{-1}(-2^{\alpha-1})$ does not exist for $\alpha>1$. 
  The 2nd TQCL keeps on moving anticlockwise with further increase of $\alpha$ till $\alpha=1.5$ beyond which it stops moving with $\alpha$ indicating the limit of the original Kitaev chain.

\subsubsection*{Case 2: When $r=3$} 
\begin{figure}[H]
 	\centering
 	\includegraphics[width=\columnwidth,height=11cm]{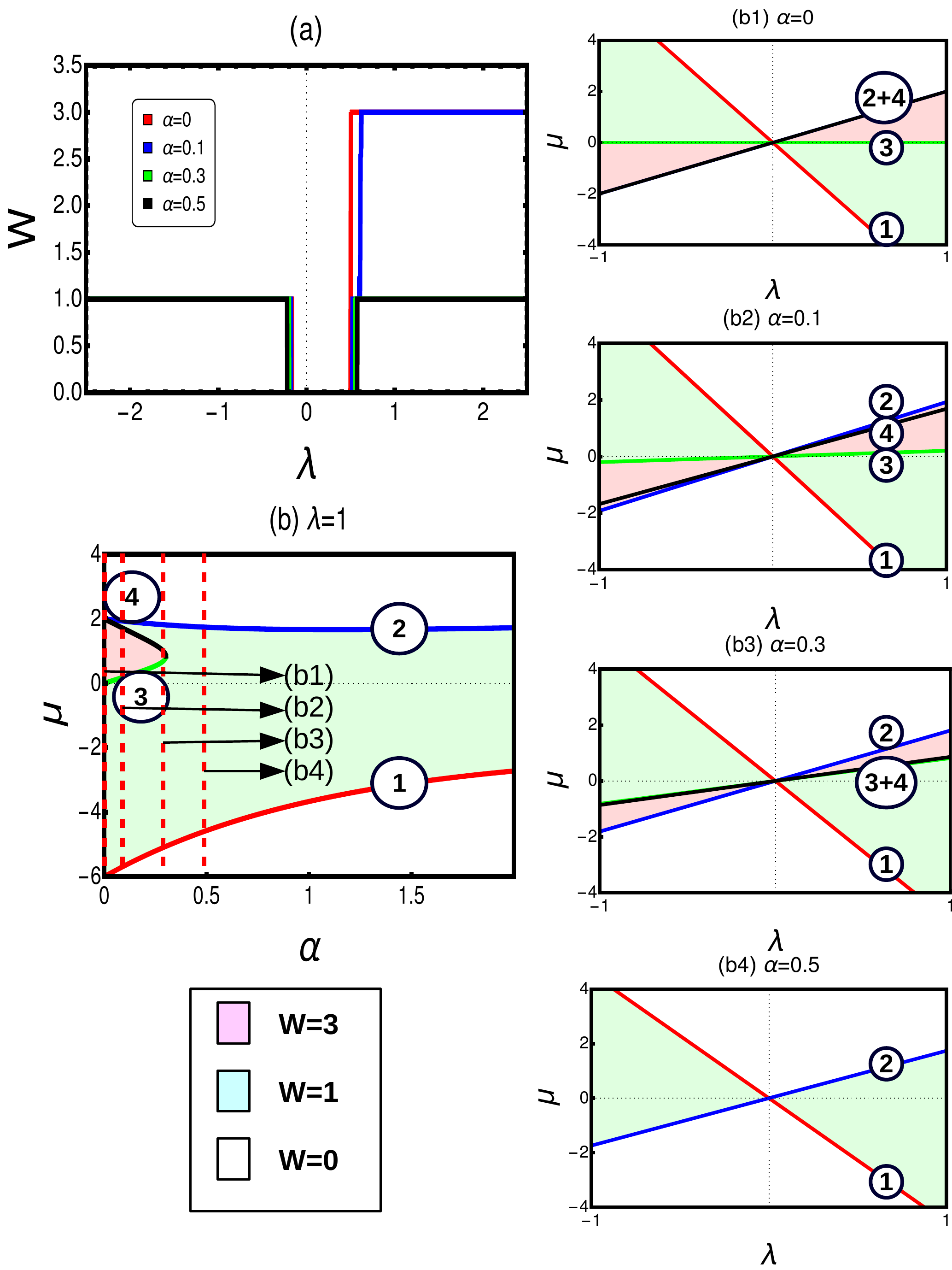}
 	\caption{Topological phase diagram of longer-range Kitaev chain with $r=3$ neighbors. a) Winding number study with respect to $\lambda$ for different values of $\alpha$. (b) Behavior of topological phases in $\mu-\alpha$ parameter space with fixed $\lambda$. (b1-b4) Corresponding topological phase diagrams of model in $\mu-\lambda$ parameter space with increasing values of $\alpha$. This helps to understand the behavior of TQCLs. Red, blue, green and black lines represent the 1st,2nd, 3rd and 4th TQCLs respectively. (b1-b2) Initially the 2nd and 4th TQCLs superpose on each other and gradually 2nd TQCL shifts upward with increasing value of $\alpha$. (b3-b4) 3rd and 4th lines superpose on each other and vanish with the increasing value of $\alpha$.}
 	\label{ff1}
 \end{figure}

Fig.~\ref{ff1}(a) shows the phase diagram for
$r=3$ where we can see the interaction is up to the 3rd nearest neighbor in the chain. Here gap closings occur at four different 
values of $k$, (Appendix~\ref{appC}) which corresponds to four different 
TQCLs.
Here the Hamiltonian is given by Eq~\ref{eq},with $J_0=\Delta_0=\lambda$ and $\alpha=\beta$.  
 For this case
\begin{equation}
W=\frac{1}{2\pi}\int_{-\pi}^{\pi}\frac{\left( 
\frac{A}{Q^2}+\frac{B}{Q}\right) }{\left( \frac{4 \lambda 
^2 P^2}{Q}+1\right) }dk,\label{WN2}
\end{equation}
where
\begin{eqnarray}
P&=&\left(2^{-\alpha } \sin (2 k)+3^{-\alpha } \sin 
(3 k)+\sin (k)\right),\nonumber\\ Q&=&\left(-2 \lambda  
\left(2^{-\alpha } \cos (2 k)+3^{-\alpha } \cos (3 
k)+\cos (k)\right)-\mu \right),\nonumber\\
A&=&4 \lambda ^2 P \left(-2^{1-\alpha } \sin (2 
k)-3^{1-\alpha } \sin (3 k)-\sin (k)\right),\nonumber\\ 
B&=&2 \lambda  \left(2^{1-\alpha } \cos (2 
k)+3^{1-\alpha } \cos (3 k)+\cos (k)\right).\nonumber
\end{eqnarray}
Here the model should contain four topologically distinct phases i.e., $W=0,1,2$ and 3 respectively. However, we observe the suppression of $W=2$ region. For $\alpha=0,0.1$, there is transition among $W:3\leftrightarrow0$ and 
 $W:1\leftrightarrow0$. 
 For $\alpha\geq0.3$, we see only $W=1$ and $W=0$ phases.
 When $\alpha=0$, the 2nd and 4th critical lines superpose on each other and results in the suppression of $W=2$ phase. But when we gradually increase the value of $\alpha$, the 3rd and 4th critical lines vanish as shown in Figs.~\ref{ff1} (b1-b4). So throughout this precess, $W=2$ region is absent. Hence the model transforms to original Kitaev model.
 \subsubsection*{Case 3: When $r=4$}  
\begin{widetext}

\begin{figure}[H]
     	\centering
     	\includegraphics[width=16cm,height=13cm]{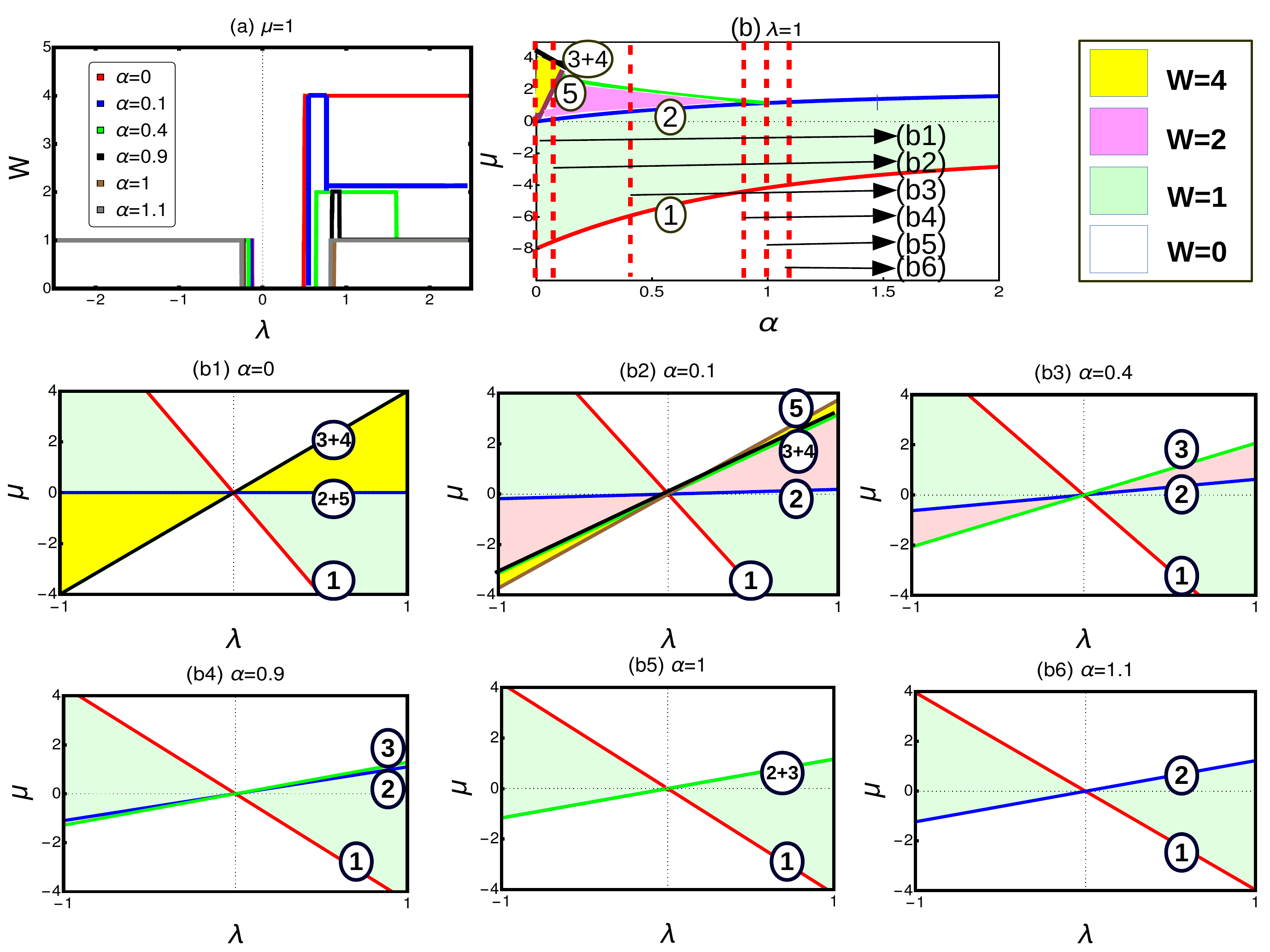}
     	\caption{Topological phase diagram of longer-range Kitaev chain with $r=4$ neighbors. a) Winding number study with respect to $\lambda$ for different values of $\alpha$. (b) Behavior of topological phases in $\mu-\alpha$ parameter space with fixed $\lambda$. (b1-b6) Corresponding topological phase diagrams of model in $\mu-\lambda$ parameter space with increasing values of $\alpha$. This helps to understand the behavior of TQCLs. Red, blue, green, black and brown lines represent the 1st, 2nd, 3rd, 4th and 5th TQCLs respectively. Initially the 2nd TQCL lies on $\lambda$-axis superposed with and gradually moves upward with increasing $\alpha$. (b1) There is superposition among 3rd-4th and 2nd-5th TQCLs. (b2) 5th TQCL shifts upward and 3rd-4th TQCLs remain as before. (b3-b4) With the increasing $\alpha$, 4th and 5th TQCLs vanish by yielding imaginary values. (b5) The 3rd TQCL superposes with 2nd TQCL. (b6) 3rd TQCL vanishes and model reduces to short-range Kitaev chain.}
     	\label{ff2}
     \end{figure}

\end{widetext}

Fig.~\ref{ff2}(a) shows the phases for $r=4$. Here we can see the interaction is up to the 4th nearest neighbor in the chain and the gap closing occurs at five different 
  values of $k$ (Appendix~\ref{appC}) 
  which corresponds to five different TQCL. 
     Technically the model should contain five topologically distinct phases i.e., $W=0,1,2,3$ and 4 respectively. 
For this case, WN is given by
\begin{equation}
W=\frac{1}{2\pi}\int_{-\pi}^{\pi}\frac{\left( \frac{P}{B^2}+\frac{Q}{B}\right) }
{\left( \frac{4 \lambda^2 A^2}{B^2}+1\right) }dk,\label{WN3}
\end{equation}
 where 
\begin{eqnarray}
A&=&\left( \frac{\sin (2 k)}{2^{\alpha }}+\frac{ 
\sin (3 k)}{3^{\alpha }}+\frac{ \sin (4 k)}{4^{\alpha 
}}+\sin (k)\right),\nonumber\\
B&=&-2 \lambda  \left(\frac{ \cos (2 
k)}{2^{\alpha }}+\frac{ \cos (3 k)}{3^{\alpha }}+ 
\frac{\cos (4 k)}{4^{\alpha }}+\cos (k)\right)-\mu,\nonumber\\
P&=&4 \lambda ^2 A \left(- \frac{\sin (2 k)}{2^{-1+\alpha 
}}- \frac{\sin (3 k)}{3^{-1+\alpha }}- \frac{\sin (4 
k)}{4^{1-\alpha }}-\sin (k)\right),\nonumber\\
Q&=&2 \lambda  \left( \frac{\cos (2 k)}{2^{-1+\alpha 
}}+ \frac{\cos (3 k)}{3^{-1+\alpha }}+ \frac{\cos (4 
k)}{4^{-1+\alpha }}+\cos (k)\right).\nonumber
\end{eqnarray}
When $\alpha=0$, we observe the transition among $W:4\leftrightarrow0$ and $W:1\leftrightarrow0$. When $\alpha=0,0.1$, there is a
  transition among $W:0\leftrightarrow4$ and $4\leftrightarrow2$. When $\alpha=0.4,0.9$, there occurs a 
  transition among $W:2\leftrightarrow1$, $2\leftrightarrow0$
  and $W:1\leftrightarrow0$. When $\alpha=1,1.1$, there is a transition only among $W:1\leftrightarrow0$, 
  which represents original Kitaev chain.
   For $\alpha=0,0.1$, there is a superposition among 3rd and 4th critical lines which results in the suppression of $W=3$ and $W=2$ phases. In the beginning, the 2nd critical line lies on the $\lambda$ axis along with fifth critical line and gradually shifts upward with the increase of $\alpha$ (Fig~\ref{ff2}(b1-b6)). This results in the formation of $W=2$ region. The 4th critical line fails to distinguish the topological phases and the 5th critical line vanishes for $\alpha>0.1$, thus throughout the process $W=3$ phase is suppressed. Hence once again the model shifts to original Kitaev chain. \\

From this section one can summarize the observations as following:
\begin{itemize}
\item We notice that by increasing the number of interacting neighbors $r$ it is possible to get higher order WNs whereas if we increase the decay parameter $\alpha$ the higher order WNs vanish.
\item The reduction of higher order WNs to lower order occurs through the process of superposition and vanishing of TQCLs.
\item Higher order WNs are less stable (decay early) compared to its lower orders. In the same way, the TQCLs associated with higher order WN are also less stable and undergo superposition/vanishing early compared to its lower orders.
\item We also notice that it is not possible to achieve all the intermediate higher order WNs. This is because some of the TQCLs responsible for those particular intermediate WN undergoes superposition and results in the suppression of corresponding topological phases.
\end{itemize}
 
\section{Characteristic study of critical lines}\label{char}
The conclusion drawn from the previous section  leads to further analysis of the TQCLs as a function of $\alpha$.
Our goal is uncovering the nature of the resultant TQCL when there exists multi-criticality i.e., when two TQCLs of distinct natures superpose on each other.
Also we would like to analyze the stability of the different TQCLs as $\alpha$ varies.
To understand these factors, we analyze the Berry connection and ground state energy of the system in this section.
\subsection{Superposition of critical lines: an analysis of Berry Connection}
Now we present the physical explanations of superposition of TQCLs from the perspective of curvature function.
Curvature function of Bloch state is an important 
quantity whose integral over the Brillouin zone defines 
the topological invariant~\cite{molignini2020unifying,chen2019topological}. The curvature function 
can take various forms like Berry connection, Berry 
curvature and Pfaffian of Bloch states. Here we consider Berry connection as our curvature 
function $F(k,M)$.

\begin{figure}[H]
	\centering
	\includegraphics[width=\columnwidth,height=12cm]{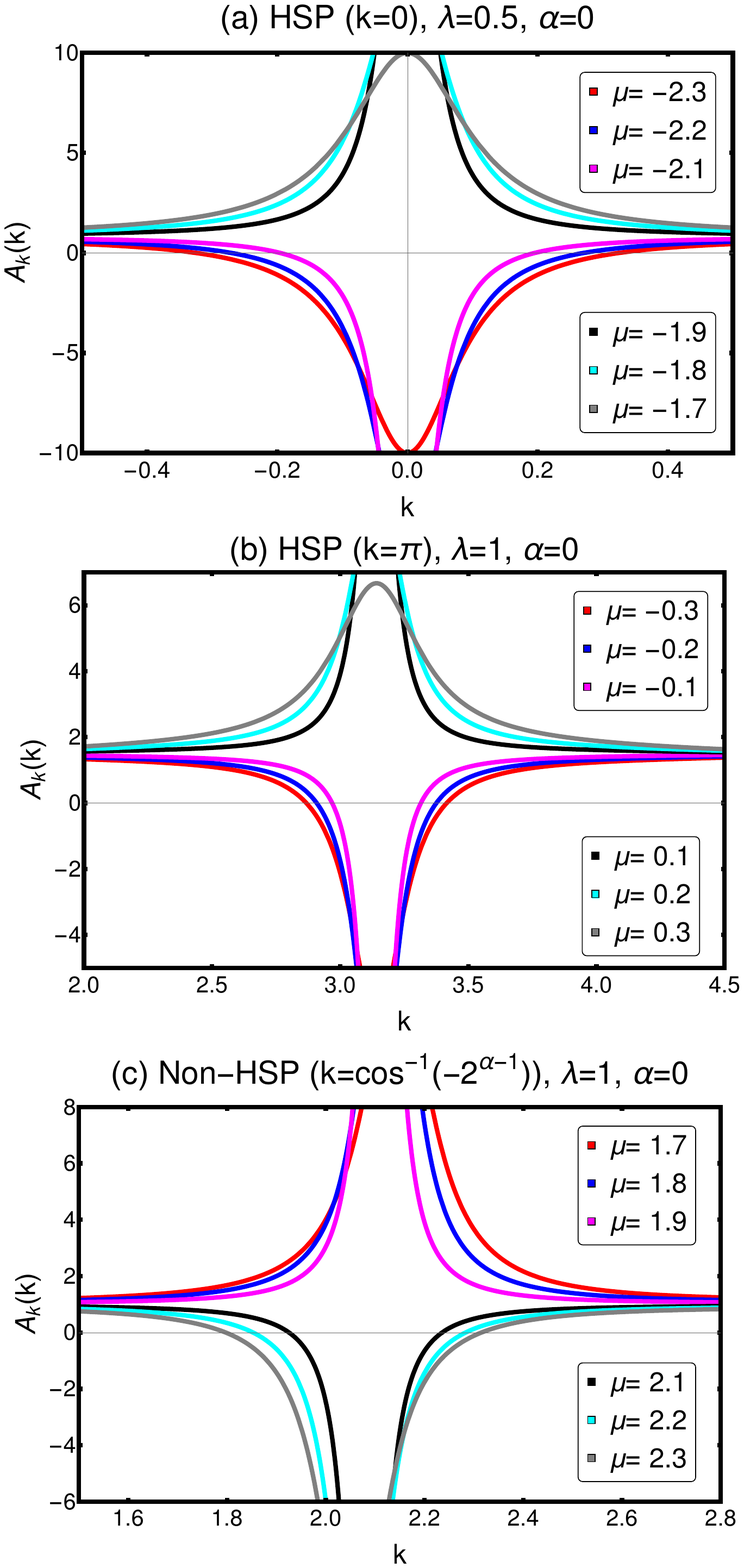}
	\caption{Behavior of Berry connection for $r=2$ 
	for (a) 1st TQCL with HSP at $k=0$ (b) 2nd TQCL with a HSP at $k=\pi$, (c) 3rd TQCL with a non-HSP $k=\cos^{-1}(-2^{\alpha-1})$. }
	\label{5}
\end{figure}

Berry connection (BC) is a momentum dependent function which diverges at specific points in the Brillouin zone as one approaches the  
critical values in the parameter space ($\mathbf{M}\rightarrow \mathbf{M_c}$, where $\mathbf{M}$ is the set of all parameters).
If these points in the Brillouin zone have a symmetry $k_0=-k_0$, then they are called as the
high symmetry points (HSP) (Fig.~\ref{5} (a,b)). 
Usually the BC behaves as an even function (i.e., $F(k_0+\delta k,M)=F(k_0-\delta 
k,M)$) around such points.
There are also points in the Brillouin zone, where the symmetry $k_0=-k_0$ is not obeyed. 
These are non-HSPs which lies in the Brillouin zone other than $k=0$ and $k=\pi$. As the parameters approaches critical value $\mathbf{M}\rightarrow\mathbf{M_c}$, the diverging peak of BC shifts towards non-HSP (Fig.~\ref{5}(c)).
In both HSP as well as non-HSP the BC diverges as one approaches the critical point ($\mathbf{M}\rightarrow\mathbf{M_c}$). As the 
critical point is crossed, the BC flips its sign, but 
point of divergence in $k$-space remains same. This is the generic nature of 
HSP. In the same way, even for a non-HSP, the BC tends to 
diverge as one approaches critical point.  
The point of divergence shifts based on the parameter 
space. This is the behavior of non-HSP.\\
Around all the critical points, the BC shows non-analytic behavior and acquires the Ornstein-Zernike form around HSPs \cite{chen2019topological} i.e.,
\begin{equation}
F(k_0+\delta k,\mathbf{M})=\frac{F(k_0,\mathbf{M})}{1+\xi_{k_0}^2\delta k^2}.\label{lor}
\end{equation}
When $k=k_0$ and with $\mathbf{M}\rightarrow\mathbf{M_c}$ the length scale diverges ($\xi_{k_0}\rightarrow\infty$), which results in the narrowing of Lorentzian term of Eq.~\ref{lor}. Thus one can observe the divergence of BC as $\mathbf{M}$ is approached from both sides of critical line~\cite{chen2019topological} i.e.,
\begin{equation}
\lim_{\mathbf{M}\rightarrow \mathbf{M}_c^+}F(k_0,\mathbf{M})=-\lim_{\mathbf{M}\rightarrow \mathbf{M^}_c^-}F(k_0,\mathbf{M})=\pm\infty.
\end{equation} 
Close to the critical point, the BC follows the relation
\begin{equation}
F(k_0,\mathbf{M})\propto|\mathbf{M}-\mathbf{M}_c|^{-\gamma} \hspace{0.25cm}\textit{and}\hspace{0.25cm} \xi(k_0,\mathbf{M})\propto|\mathbf{M}-\mathbf{M}_c|^{-\nu},\label{exp}
\end{equation}
where the exponents $\gamma$ and $\nu$ corresponds to the susceptibility and localization critical exponents respectively. For a one dimensional model, exponents obey the scaling law $\gamma=\nu$ and take modified forms near multi-critical points\cite{kumar2020multi}.\\
Here we try to analyze the behavior of 
TQCLs with the decay parameter $\alpha$. We 
consider some cases for different values of $r$ and analyze the BC to understand the behavior of TQCLs 
when they undergo the process of superposition. The
BC for a Bloch state across the 
Brillouin zone is defined as,
\begin{eqnarray}
A_k=-2i\left\langle 
u_{nk}|i\partial_k|u_{nk}\right\rangle=\frac{\chi_z\partial_k
\chi_y-\chi_y\partial_k\chi_z}{\chi_z^2+\chi_y^2},
\label{cf1}
\end{eqnarray}
where $|u_{nk}\rangle= 
\frac{1}{\sqrt{2}\chi}\left( \begin{matrix}
-\chi\\\chi_z+i\chi_y
\end{matrix}\right) $. For example, when $r=2$,
\begin{eqnarray}
A_k=\frac{2 \lambda  \left(2 \left(4^{\alpha }+2\right) 
\lambda +2^{\alpha }A\right)}{4 \lambda 
^2+4^{\alpha }B+2^{\alpha +2} \lambda  (2 
\lambda  \cos (k)+\mu  \cos (2 k))}.
\end{eqnarray}
where 
\begin{eqnarray}
A&=&\cos (k) \left(2^{\alpha } \mu +6 \lambda 
\right)+2 \mu  \cos (2 k),\nonumber\\
B&=&4 \lambda ^2+4 \lambda  
\mu  \cos (k)+\mu ^2.\nonumber
\end{eqnarray}
Here HSPs are $k_0=0,\pi$ and 
non-HSP is $k_0=\cos^{-1}(-2^{\alpha-1})$. As there are 
just two interacting neighbors, we have three 
critical points (Appendix~\ref{appC}).

In a longer-range Kitaev model there always exists two HSPs at $k=0$ and $k=\pi$. The number of non-HSPs depends on the number of interacting neighbors $r$ and higher order $r$ generates higher order TQCLs. When the decay parameter $\alpha$ starts to increase, the higher order TQCLs start to superpose on each other and vanish. Here we analyze different kind of superpositions:
superposition of two HSP,
superposition of HSP and non-HSP and
superposition of two non-HSP.
We study above combinations in the following.
\subsubsection*{Case 1: When $r=2$} 
 \begin{figure}[H]
  	\centering
  	\includegraphics[width=7cm,height=10cm]{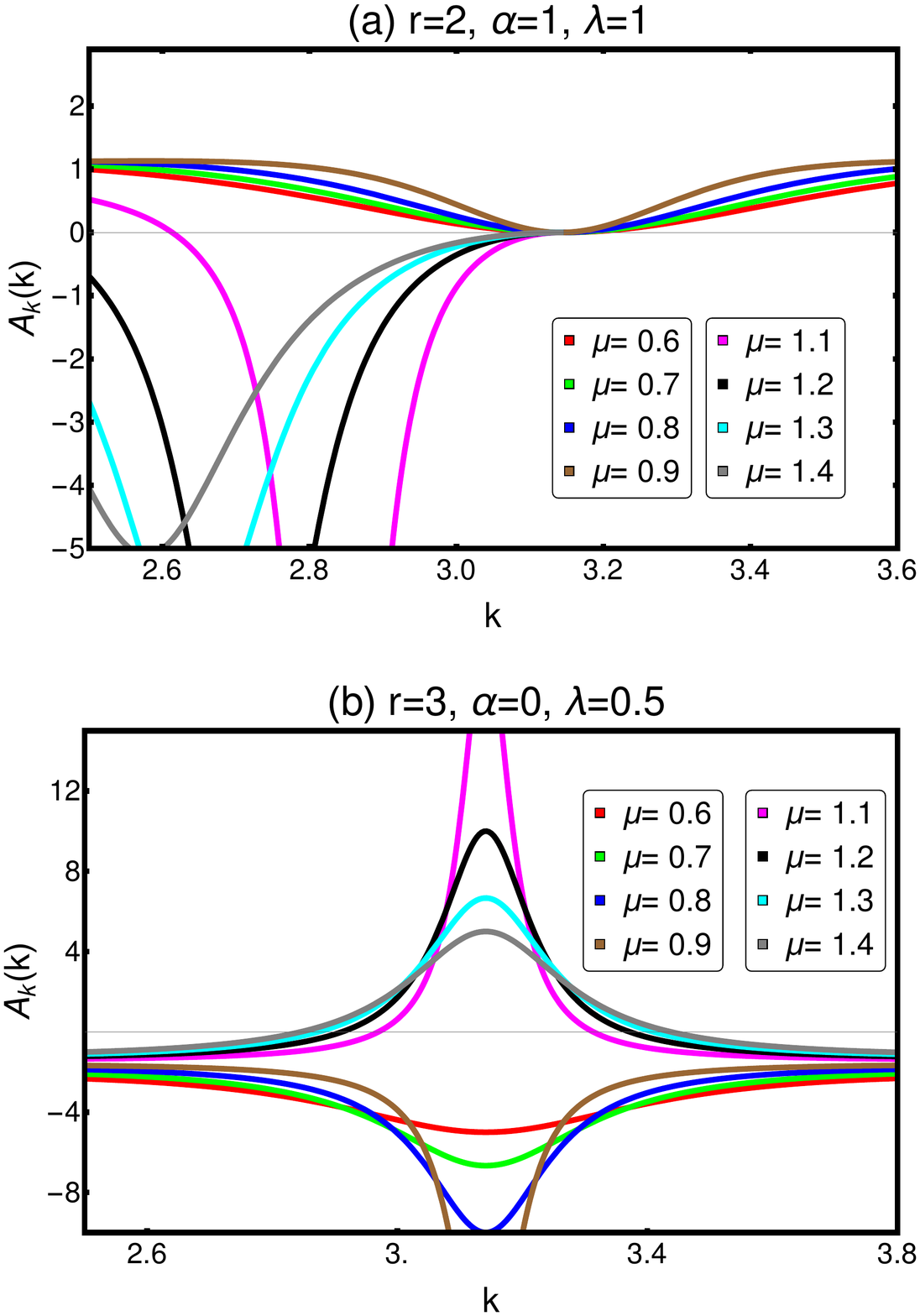}
  	\caption{Behavior of Berry connection 
  	for  
  	a) When $r=2$, superposition of 2nd TQCL (HSP) and 3rd TQCL (non-HSP). b) When $r=3$, superposition of 2nd (HSP) and 4th TQCLs (non-HSP).}
  	\label{6}
  \end{figure}
 Fig.~\ref{6}(a) shows the superposition of 2nd (HSP) 
and 3rd (non-HSP) TQCL for $\alpha=1$ and $\lambda=1$. 
The 2nd TQCL starts moving 
in the anticlockwise direction with increase in $\alpha$ (see Fig.~\ref{r2pd}). When $\alpha=0$ we can 
see the symmetric behavior of HSP at $k=\pi$. The 
BC tends to diverge at the point ($\mu=0,\alpha=0,\lambda=1$) which happens to lie on the 2nd TQCL (Fig.~\ref{5}(b)).
As critical 
point ($\mu=2,\alpha=0,\lambda=1$) on 3rd TQCL is crossed , the BC 
flips its sign, but the point of divergence in $k$-space is not same (Fig.~\ref{5}(c)).
One can observe that the behavior of the plots in Fig.~\ref{6}(a) is a combination of both HSP as well as non-HSP. As one 
approaches the critical point from the lower values of 
$\mu$ the BC behaves similar to that for a HSP i.e., it shows the peak around $k=\pi$ without any shift. But after the  critical 
point ($\mu=1,\alpha=1,\lambda=1$), it behaves as that for a
non-HSP i.e, the diverging peaks of BC starts shifting. This is an interesting phenomenon that occurs as a consequence of superposition of TQCLs.
 \subsubsection*{Case 2: When $r=3$}
In Fig.~\ref{6}(b) we observe the superposition of 2nd TQCL (HSP) and 4th TQCL (non-HSP) for $r=3$. Here we observe the symmetric nature of BC around the QCP ($\mu=1,\alpha=0,\lambda=0.5$). As we approach the QCP from the lower values of $\alpha$, we observe the evenness of BC because of the high symmetric nature of 2nd TQCL. As we pass the QCP we observe the flip in the BC, but still it exhibits high symmetric nature even though it is a non-HSP. We observe that the TQCLs associated with the higher order WNs are usually non-HSPs and are less stable with respect to $\alpha$. When they superpose with HSP, there occurs a very slight and insignificant shift in the diverging peaks, and hence they behave similar to HSP. In other way, the  HSP dominates over the non-HSP which is associated with higher order TQCLs.
\subsubsection*{Case 3: When $r=4$}
\begin{figure}[H]
	\centering
	\includegraphics[width=\columnwidth,height=14cm]{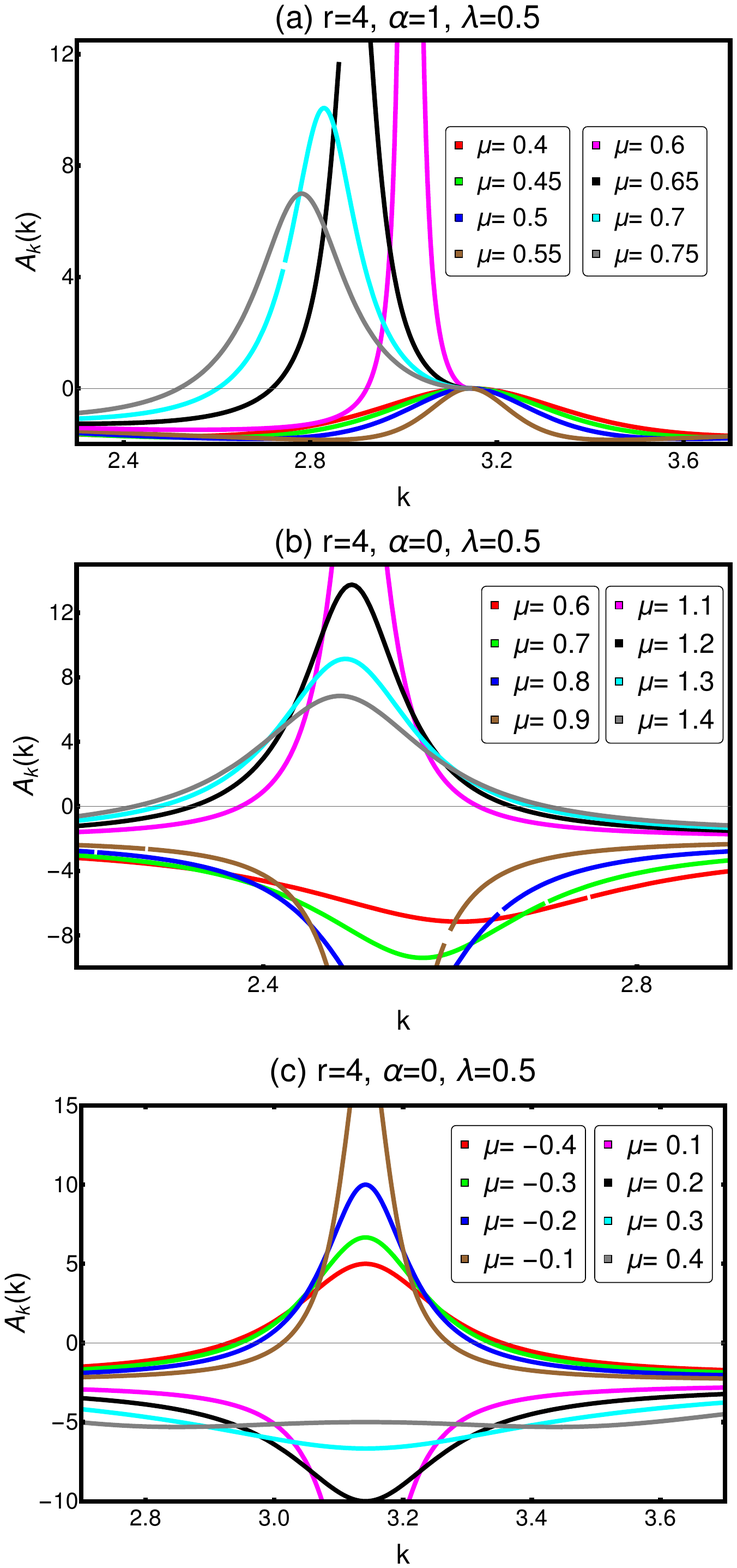}
	\caption{Behavior of Berry connection 
	for a) When $r=4$, superposition of 2nd TQCL (HSP) and 3rd TQCL (non-HSP).
	(b) When $r=4$, superposition of 
	3rd (non-HSP) and 4th (non-HSP) TQCLs. 
	 c) When $r=4$, superposition of 2nd TQCL (HSP) and 5th TQCL (non-HSP).}
	\label{7}
\end{figure}

In Fig.~\ref{7}(a) we show the superposition 2nd (HSP) and 3rd (non-HSP) TQCLs  for $r=4$.  As one approaches QCP ($\mu=0.58,\alpha=1,\lambda=0.5$) from lower values of $\mu$ it exhibits the nature of HSP and as the QCP passes it exhibits the nature of non-HSP. Here we can observe the nature of both HSP as well as non-HSP.
Fig.~\ref{7}(b) shows the superposition of 3rd and 4th TQCLs  for $r=4$ where both of them are non-HSP. As one approaches QCP ($\mu=1,\alpha=0,\lambda=0.5$) from the lower values of $\mu$, the BC shows the non-HSP nature because of the effect of 3rd TQCL. As the QCP is passed, the BC flips its sign, and still continue to behave as non-HSP. But we observe a comparatively less shift in the diverging peak of BC.

In Fig.~\ref{7}(c) we observe the superposition 2nd (HSP) and 5th (non-HSP) TQCLs  for $r=4$.  As one approaches QCP ($\mu=0,\alpha=0,\lambda=0.5$) from lower values of $\mu$ it exhibits the nature of HSP and as the QCP passes, even then it exhibits the nature similar to HSP. Here the non-HSP nature is very less expressive and one can observe an insignificant shift of diverging BC peaks. Hence the HSP dominates over non-HSP of higher order TQCLs similar to Fig.~\ref{6}(b).

The 1st TQCL (HSP $k=0$) does not involve in any superposition phenomena with the increasing values of $\alpha$.  For a short-range model, the region bounded between two HSP ($k=0,\pi$) gives $W=1$ topological phase which is also the characterizing nature of Kitaev chain. As $\alpha\rightarrow\infty$ all the longer-range models 
reduce to original Kitaev chain. Hence there is no chance for superposition of two HSP ($k=0,\pi$). If such case occurs, then there will be an absence of $W=1$ phase.
\subsubsection*{ Critical exponents for longer-range models}
Critical exponents are the quantities which explain the behavior of the system around the criticality. However, the TQPTs are second order phase transitions where one can calculate the critical exponents by expanding the pseudo-spin vectors around at $k=k_0$ as
\begin{equation}
\chi(k)|_{k=k_0}=\chi(k_0)+\chi^{\prime}(k_0)k+\chi^{\prime\prime}(k_0)k^2/2.
\end{equation}
By substituting the expanded form of pseudo-spin vectors in Eq.~\ref{cf1}, which leads to the Ornstein-Zernike form ~\cite{kumar2020multi,rufo2019multicritical}. i.e.,
\begin{eqnarray}
F(k,\mathbf{M})\mid_{k=k_0}&=&\frac{A.\delta k(2B.\delta k)-(\delta g+B\delta k^2)A}{\delta g^2+(2B\delta g+A^2)\delta k^2+B^2\delta k^4}\nonumber\\
&=&\frac{\frac{2AB\delta k^2-A(\delta g+B\delta k^2)}{\delta g^2}}{1+\left(\frac{2\delta g.B+A^2}{\delta g^2} \right)\delta k^2+\left(\frac{B^2}{\delta g^2} \right)\delta k^4  }\nonumber\\
&=&\frac{F(k_0,\delta g)}{1+\xi^2\delta k^2+\xi^4\delta k^4},
\label{cor}
\end{eqnarray}
where $\xi$ is the characteristic length. The terms $A, B$ and $\delta g$ are the parameters that come from Taylor series of expansion. In Eq.~\ref{cor}, there are two terms which decides the characteristic length. 1) $\xi\propto\sqrt{\frac{2B}{\delta g}+\frac{A^2}{\delta g^2}}$, where the term $\frac{A^2}{\delta g^2}$ dominates over $\frac{2B}{\delta g}$. Hence $\xi\propto1/|\delta g|\Rightarrow\nu=1$. 2) $\xi\propto\sqrt[4]{\frac{B^2}{\delta g^2}}$ and $\xi\propto|\delta g|^{-1/2}\Rightarrow\nu=1/2$. Thus the dominating term among $A$ and $\sqrt{B}$ decides the characteristic length critical exponent. The exponent of the numerator gives the susceptibility critical exponent \cite{kumar2020multi}. In Table~\ref{ce}, we calculate the terms $A,B$ and $\delta g$ of a longer-range Kitaev chain with different number of interacting neighbors $r$. These critical exponents only correspond to  HSPs, which always fit into scaling law whenever they are away from multi-critical points as well as superposition of TQCLs.\\
In a one dimensional system the critical exponents of Eq.~\ref{exp} follow the scaling rule $\nu=\gamma$ \cite{rufo2019multicritical}. But when there exists some multi-criticality  or superposition of critical lines, this relation does not hold good~\cite{kumar2020multi}. This is because of the unevenness of the BC around those critical point. Around the superposition of TQCL one can observe the violation of even nature of BC throughout the BZ (i.e., $lim_{\mathbf{M}\rightarrow \mathbf{M}_c^+}F(k_0,\mathbf{M}^+)\neq -lim_{\mathbf{M}\rightarrow \mathbf{M}_c^-}F(k_0,\mathbf{M}^-)=\pm\infty$ )\cite{chen2019topological}. For a non-HSP, the curvature function fails to acquire the Ornstein-Zernike form, and thus it is not possible to do conventional scaling in such cases \cite{molignini2018universal}.\\
The scaling scheme also fails at high symmetry points where the BC attains the fixed point configuration \cite{kumar2020multi}. This kind of behavior can be observed in Fig.~\ref{6}(a), \ref{7}(a). The fixed point configuration of the BC is the state where the height of the curve does not vary along with the varying parameter, which can occur as a consequence of the multi-critical point. Hence it is not possible to calculate the critical exponents $\nu$ and $\gamma$ at these points.\\\\
\noindent Through the study of BC we notice following points:
\begin{itemize}
\item Longer-range models contains a number of symmetry points (HSP/non-HSP) depending on the number of interacting neighbors.
\item As the decay parameter $\alpha\rightarrow\infty$, the longer-range model reduces to short-range where only HSP remains.
\item When two TQCLs superpose on each other, both TQCLs influence each other. Because of this reason, we can find mixed nature in the resultant TQCL.
\item The points corresponding to $k=0$ and $k=\pi$ represents HSP and all other symmetry points correspond to non-HSP. The non-HSPs corresponding to higher order WN or higher TQCL are comparatively less expressive in their nature.
\item The lower order symmetry points dominate over the higher order symmetry points when they superpose on each other.
\item It is not possible to find the critical exponents near the multi-critical points as well as at superposition of TQCLs. 
\end{itemize}

\begin{widetext}

\begin{table}[H]
 \begin{center}
\begin{tabular}{ |c|c|c|c|c| } 
\hline
\hline
Number &Critical&&&\\
of&condition&$A$&$B$&$\delta g$\\
neighbors&(HSP)&&&\\
\hline
\hline
$r=2$&$k=0$&$2\lambda(1+\frac{2}{2^{\alpha}})$&$\lambda(1+\frac{4}{2^{\alpha}})$&$-\mu-2\lambda(1+\frac{2}{2^{\alpha}})$\\
&&&&\\
&$k=\pi$&$2\lambda(-1+\frac{2}{2^{\alpha}})$&$\lambda(-1+\frac{4}{2^{\alpha}})$&$-\mu-2\lambda(-1+\frac{2}{2^{\alpha}})$\\
&&&&\\
\hline
\hline
$r=3$&$k=0$&$2\lambda(1+\frac{2}{2^{\alpha}}+\frac{3}{3^{\alpha}})$&$\lambda(1+\frac{4}{2^{\alpha}}+\frac{9}{3^{\alpha}})$&$-\mu-2\lambda(1+\frac{2}{2^{\alpha}}+\frac{3}{3^{\alpha}})$\\
&&&&\\
&$k=\pi$&$2\lambda(-1+\frac{2}{2^{\alpha}}-\frac{3}{3^{\alpha}})$&$\lambda(-1+\frac{4}{2^{\alpha}}-\frac{9}{3^{\alpha}})$&$-\mu-2\lambda(-1+\frac{2}{2^{\alpha}}-\frac{3}{3^{\alpha}})$\\

&&&&\\\hline
\hline
$r=4$&$k=0$&$2\lambda(1+\frac{2}{2^{\alpha}}+\frac{3}{3^{\alpha}}+\frac{4}{4^{\alpha}})$&$\lambda(1+\frac{4}{2^{\alpha}}+\frac{9}{3^{\alpha}}+\frac{16}{4^{\alpha}})$&$-\mu-2\lambda(1+\frac{2}{2^{\alpha}}+\frac{3}{3^{\alpha}}+\frac{4}{4^{\alpha}})$\\
&&&&\\
&$k=\pi$&$2\lambda(-1+\frac{2}{2^{\alpha}}-\frac{3}{3^{\alpha}}+\frac{4}{4^{\alpha}})$&$\lambda(-1+\frac{4}{2^{\alpha}}-\frac{9}{3^{\alpha}}+\frac{16}{4^{\alpha}})$&$-\mu-2\lambda(-1+\frac{2}{2^{\alpha}}-\frac{3}{3^{\alpha}}+\frac{4}{4^{\alpha}})$\\
&&&&\\
\hline
\hline
\end{tabular}
\end{center}
\caption{Possible parameter values of pseudo-spin vectors after series expansion. Here the $\delta g$ gives the criticality ($\mathbf{M}\rightarrow\mathbf{M_c}$) condition. The terms $A$ and $B$ decide the values of characteristic length critical exponent. In all the above cases, for $\alpha>1$ the term $A$ dominates and critical exponents yield $\gamma=\nu=1$. However, for $\alpha<1$ the values of $\nu$ and $\gamma$ varies depending on the number of interacting neighbors $r$.}
\label{ce}
\end{table}

\end{widetext}

\subsection{Vanishing of critical lines}
TQCLs are the boundaries which separates distinct topological phases and manifests as a gap closing in the quasi-energy spectrum. All the TQCLs lead to the gap closing but all the gap closings need not correspond to TQCLs. Vanishing of TQCL occurs through different means as follows,
\begin{itemize}
\item By creating the gap openings in the quasi-energy spectrum.
\item The critical line acquires complex values and at some point real part vanishes and only imaginary part remains (Complex critical lines do not have any physical meaning thus we consider only real part of such critical lines).
\end{itemize}
TQPTs are the discontinuities in the second order derivative of the ground state energy of the system~\cite{chen2008intrinsic,kempkes2016universalities,cats2018staircase}. These discontinuities represent the non-analyticities of
the ground state energy\cite{niu2012majorana,malard2020multicriticality}i.e. the ground state energy before and after TQPTs belongs to two different topological indices. This creates a discontinuity in the second order derivative of the ground state energy.
In this section, 
we study ground state energy of the system to understand the 
TQPT and vanishing of TQCL as mentioned earlier.
\subsubsection*{Case 1: When $r=2$}  
\begin{figure}[H]
 	\centering
 	\includegraphics[width=6cm,height=8cm]{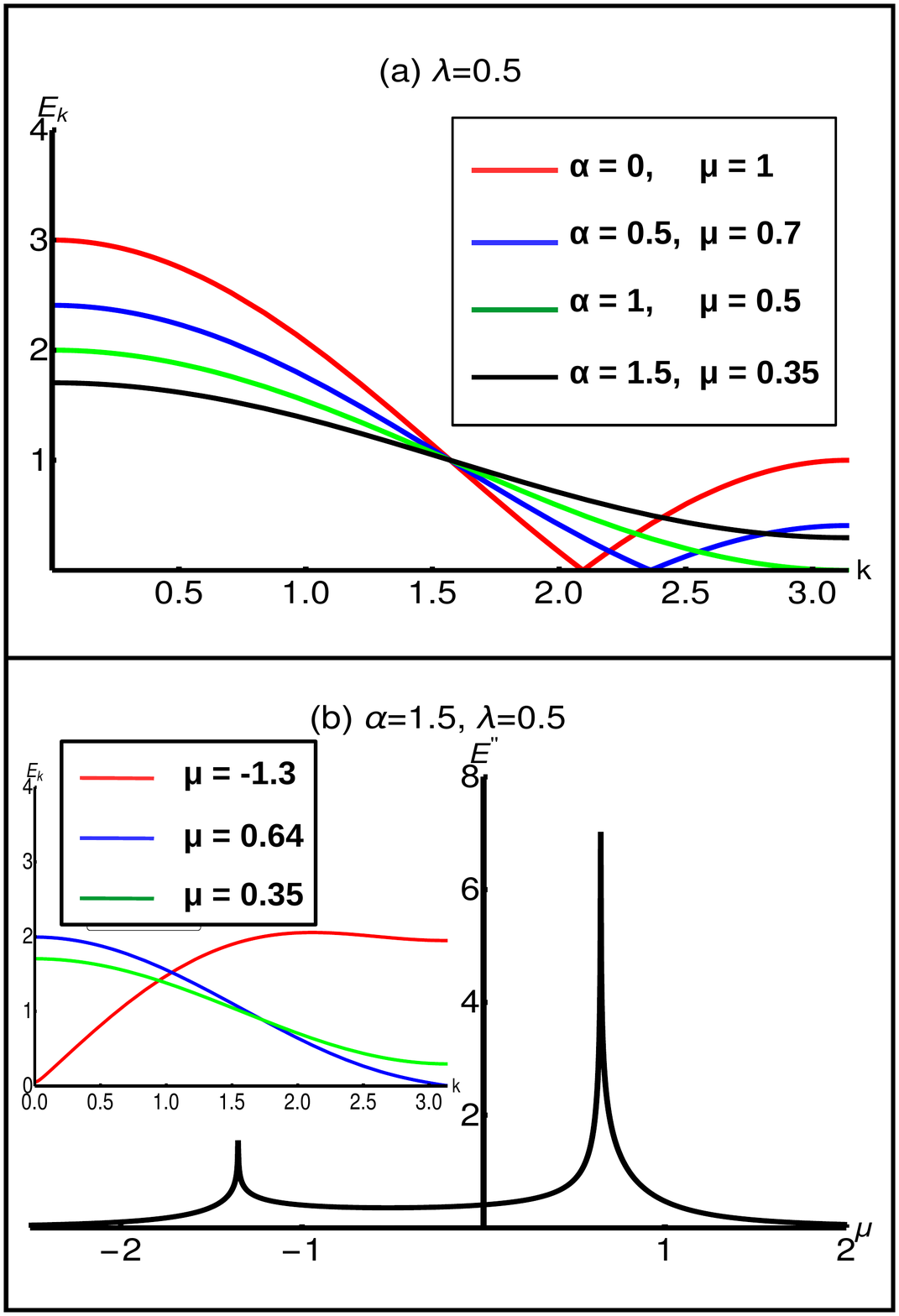}
 	\caption{
 		 The panel indicates the 
 		process of vanishing of critical line for $r=2$. a) Gap closing 
 		of 3rd critical line for different values of 
 		$\alpha$. Here the black line does not create gapless condition. b) Discontinuity in the second order 
 		derivative of ground state energy. Two spikes corresponds to 1st and 
 		2nd TQCL respectively. {\itshape Inset:} Behavior of all the three critical lines for 
 				$\alpha=1.5$. The green line (3rd TQCL) does not create gap closing.}
 	\label{gs}
 \end{figure}

From Fig.~\ref{r2pd} we observe that with the increasing values of $\alpha$, the $W=2$ phase vanishes along with its associated TQCL. To verify this we consider the 3rd TQCL and study the quasi-energy dispersion as a function of $\alpha$. One can always find a gapless point at $k=\cos^{-1}(-2^{\alpha-1})$ in the quasi-energy spectrum for $\alpha<1$. However the quasi-energy spectrum becomes gapped $\forall k$ for $\alpha>1$ as shown in Fig.~\ref{gs}(a).
Another way to verify the vanishing of TQCL is through the study of ground-state energy of the Hamiltonian. 
In topological systems a phase transition can be understood from the non-analyticities of
the ground-state energy\cite{niu2012majorana,malard2020multicriticality}. 
Here we calculate the second order derivative of 
ground-state energy $E^{''}(\mu)$, given by
\begin{eqnarray}
\frac{\partial^2E(\mu)}{\partial\mu^2}&=&-\frac{1}{2\pi}\int_{-\pi}^{\pi}\frac{\partial^2}{\partial\mu^2}
\left(
\sqrt{\left(\chi_z 
\right)^2+\left(\chi_y\right)^2}\right)dk,
\nonumber\\
&=&-\frac{1}{2\pi}\int_{-\pi}^{\pi}\frac{2 \lambda ^2 \sin ^2(k) \left(2^{\alpha }+2 \cos 
(k)\right)^2}{  \left(4 \lambda ^2+4^{\alpha }Q+2^{\alpha +2} \lambda  A\right) \sqrt{P}}dk,\nonumber
\end{eqnarray}
where
\begin{eqnarray}
P&=&4^{1-\alpha } \lambda  \left(4^{\alpha } (\mu  
\cos (k)+\lambda )+2^{\alpha } A+\lambda \right)+\mu 
^2,\nonumber\\
Q&=&4 
\lambda ^2+4 \lambda  \mu  \cos (k)+\mu 
^2,  \nonumber\\
A&=&(2 \lambda  \cos (k)+\mu  \cos (2 k)).\nonumber
\end{eqnarray}

Fig.~\ref{gs}(b) shows the discontinuities (spikes) in $E^{''}(\mu)$ for $\lambda=0.5$, $\alpha=1.5$ and for two different values of $\mu$ which correspond to 1st and 2nd TQCLs respectively. We do not observe the third spike which is the signature for 3rd TQCL. This means for 
$\alpha>1$, 3rd TQCL vanishes (see Appendix~\ref{appC}). When one increases the number of interacting 
neighbors $r$, the number of different topological phases increases leading to
more number of TQCLs. As decay parameter 
$\alpha\rightarrow\infty$, gradually all higher order WNs along with TQCLs (corresponding to non-HSPs) vanish. At the end, only the TQCL which 
corresponds to HSP ($k=0, \pi$) remains. This remaining TQCL characterizes the TQPT between $W=0$ and $W=1$ phases in the original Kitaev chain.
 One can verify this by the quasi-energy spectrum for $\alpha>1$. In the inset of Fig.~\ref{gs}(b)  we study all three TQCL for $\alpha=1.5$ $(\alpha>1)$. Here we can observe the gap closings only for $k=0$ and $k=\pi$ which correspond to the 1st and 2nd TQCLs respectively. There does not exist a third gap closing point. Hence it signals the absence of 3rd TQCL for $\alpha>1$.
 
 \subsubsection*{Case 2: When $r=3$}
\begin{figure}[H]
   	\centering
   	\includegraphics[width=\columnwidth,height=8cm]{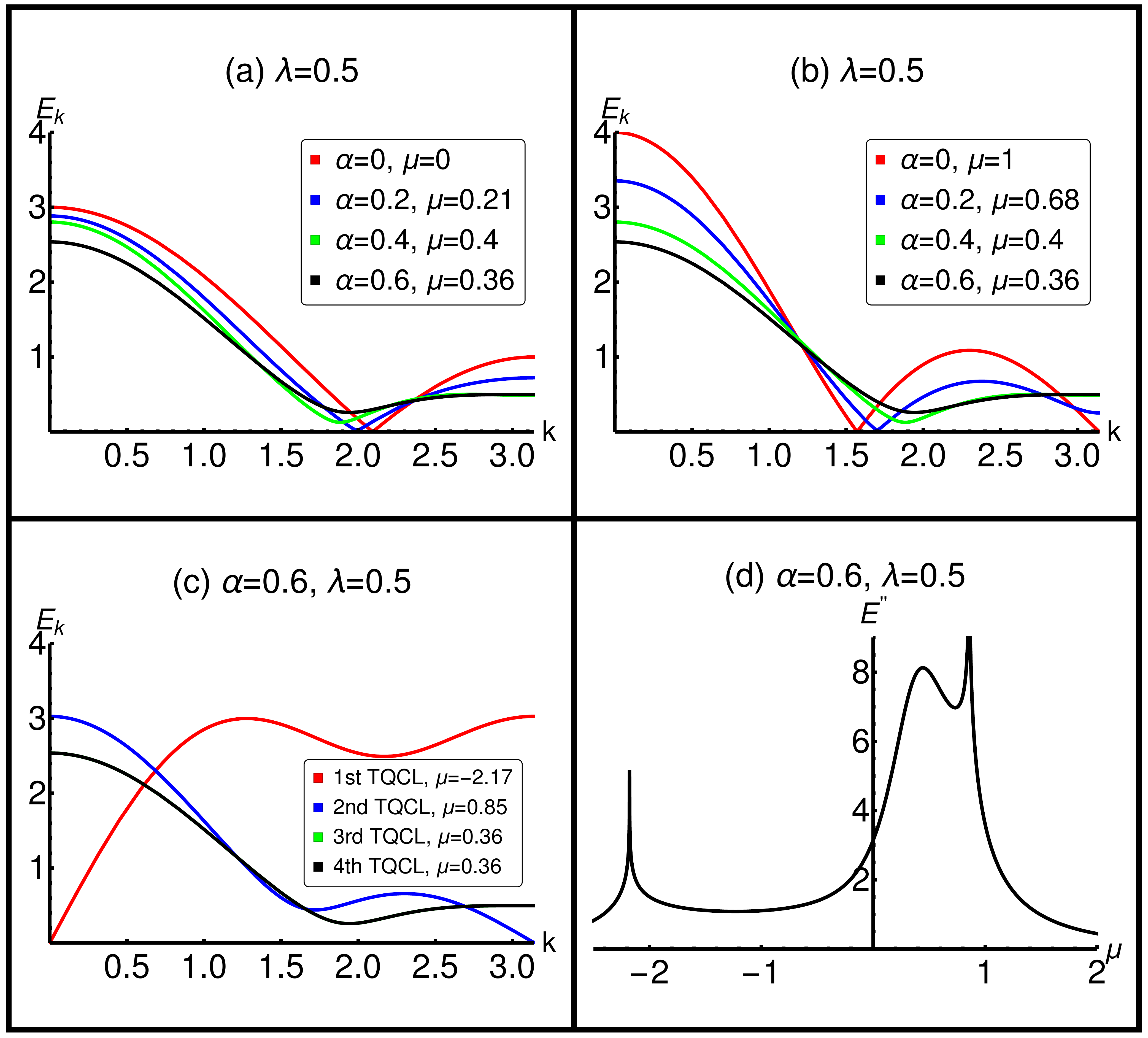}
   	\caption{
   		 The panel indicates the 
   		process of vanishing of critical line $r=3$. a) Gap closing 
   		of 3rd TQCL for different values of 
   		$\alpha$. Here the green and black lines does not create gapless condition. b) Gap closing 
   			of 4th TQCL for different values of 
   			$\alpha$. Even here, the green and black lines does not create gapless condition. c) Behavior of all the four TQCLs for 
   				$\alpha=0.6$. The green and black lines does not create gap closing. d) Discontinuity in the second order 
   						derivative of ground state energy. Two spikes corresponds to 1st and 
   						2nd TQCL respectively. There is no spike for 3rd and 4th TQCL.}
   	\label{van}
   \end{figure}

  Fig.~\ref{van} represents the process of vanishing of TQCL for $r=3$. The system consists of four TQCLs among which the first two are HSP and later two are non-HSP respectively. Initially, at $\alpha=0$ the energy spectrum corresponding to 3rd TQCL is gapless and becomes gapped after $\alpha=0.2$ (Fig. \ref{van}(a)). In the same way the 4th TQCL also exhibits a transformation from gapless to gapped spectrum (Fig. \ref{van}(b)). In both the cases the gapped energy spectrum represents the vanishing of TQCL. At the end of process one can observe only two gapless TQCLs which correspond to HSP i.e., $k=0$ and $\pi$ respectively (Fig. \ref{van}(c)).
   
   This can also be verified by the second order derivative of the ground-state energy, given by,\\
\begin{eqnarray}
\frac{\partial^2E(\mu)}{\partial\mu^2}&=&-\frac{1}{2\pi}\int_{-\pi}^{\pi}\frac{2 \lambda ^2 P^2}{\pi  \left(4 \lambda ^2 P^2+(\mu +2 \lambda  Q)^2\right)^{3/2}}dk,\nonumber
\end{eqnarray}
where
\begin{eqnarray}
P&=&2^{-\alpha } \sin (2 k)+3^{-\alpha } \sin (3 k)+\sin (k),\nonumber\\
Q&=&2^{-\alpha } \cos (2 k)+3^{-\alpha } \cos (3 k)+\cos (k).\nonumber
\end{eqnarray}
For $r=3$, the term $E^{\prime\prime}(k)$ shows only two  non-analyticities for $\alpha>0.2$ which correspond to HSPs (Fig. \ref{van})(d)). The small hump before the second spike represents former QCP. With the increasing value of $\alpha$, the discontinuities in $E^{\prime\prime}$ (which correspond to 3rd and 4th TQCL) vanish and  gradually become a continuous curve. This signals the vanishing of 3rd and 4th TQCL for higher values of $\alpha$.
 \subsubsection*{Case 3: When $r=4$} As one goes for more number of interacting neighbors $r$, it is possible to witness complex TQCLs (see Appendix~\ref{appC} for details). As per present knowledge, complex TQCLs does not have any particular physical significance. In $r=4$ case, initially for lower values of $\alpha$ the 3rd, 4th and 5th TQCL participates in the process of superposition of TQCLs. But as $\alpha$ increases, the TQCLs becomes complex and gradually the real part becomes zero. The imaginary part does not contribute to the phase boundary of topological phases. Hence one can come to conclusion that, the occurrence of imaginary value is the signature of vanishing of TQCLs. 
\subsection{Longer-range effect and the stability of higher order winding numbers}\label{stability}
\begin{figure}[H]
			\centering
			\includegraphics[width=\columnwidth,height=14cm]{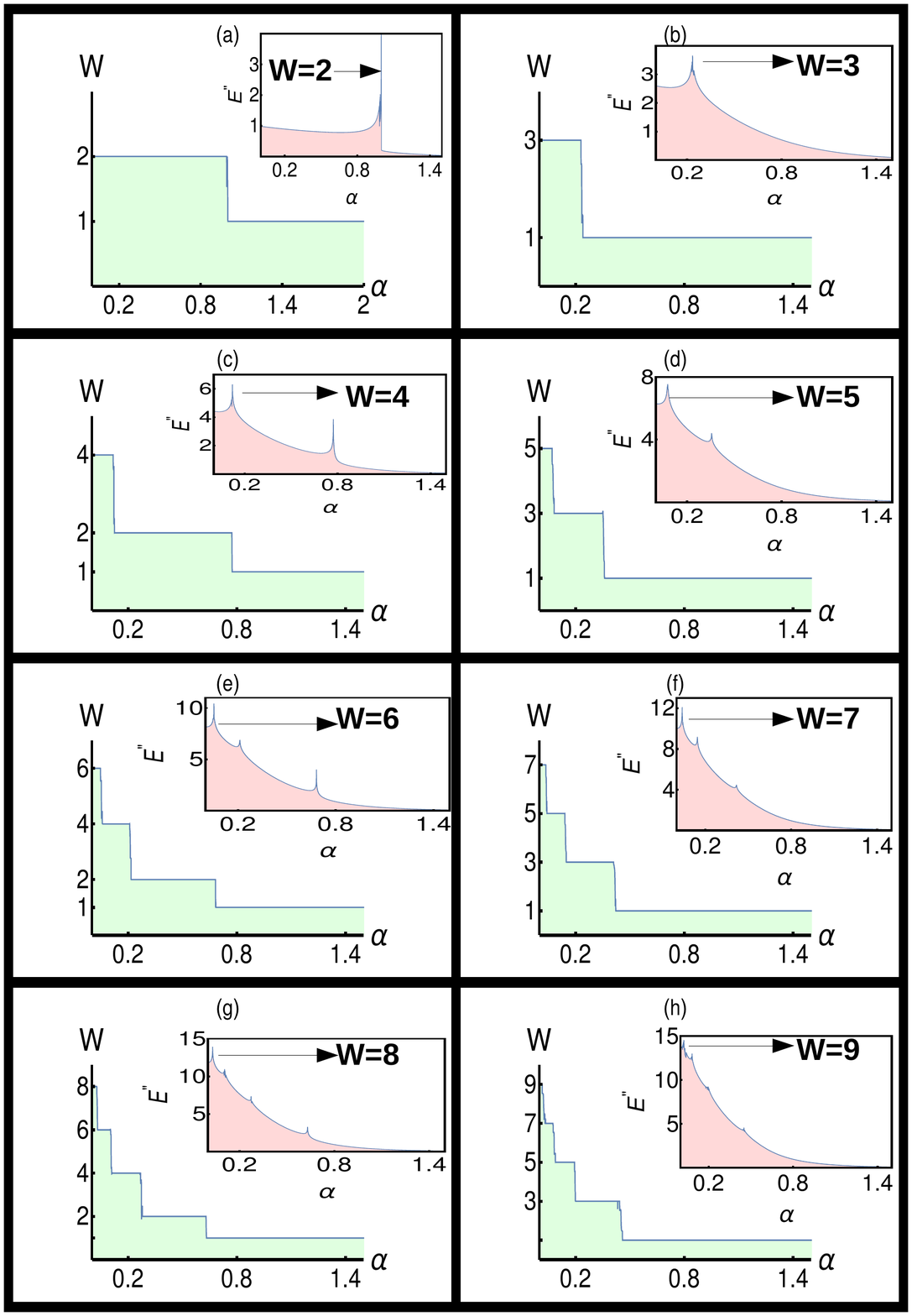}
			\caption{Variation of winding 
			number with $\alpha$ for the increasing numbers of interacting nearest neighbors $r=2,3,4,...,9$ as shown in figure (a), (b), (c),..., (h) respectively. 
			The TQPT from higher WN to lower occurs among even-even or odd-odd WNs only.
			{\it Inset:} The corresponding second order derivative of the ground state energy $(E^"=\frac{\partial^2 E}{\partial \alpha^2})$ as a function of $\alpha$. The peaks in the plots denote the points of TQPTs. For all the plots we keep $\lambda=1$ and $\mu=1$.}
			\label{avsm}
		\end{figure}

From Fig.~\ref{r2pd},\ref{ff1},\ref{ff2} we understand certain behavior of the isotropic longer-range Kitaev chain. Here we study the fate of the topological phase with highest WN corresponding to a specific value of $r$ as the decay parameter $\alpha$ is varied from $0$ to higher values. For better understanding we consider number of interacting neighbors $r>4$ as shown in Fig.~\ref{avsm} .
		We observe an interesting 
		behavior of TQPT for the values $r>2$. There exist TQPTs from even-to-even and 
		odd-to-odd WNs only for even and odd values of $r$ respectively.\\
It is well known that there exists a one-to-one correspondence between the WN as well as the localized edge modes of the system \cite{niu2012majorana,verresen2018topology}. For a system with $r$ edge modes there always exists different penetration lengths corresponding to each edge mode \cite{continentino2020finite}. At TQPT this penetration length diverges and edge mode merges with the bulk. For a system with higher WN, due to the longer-range effect as one increases $\alpha$ the edge mode decays in a faster way. This creates an instability in such a way that the higher order WN reduces to its corresponding lower order.
				 However, in the large $\alpha$ (short-range) regime all higher order WNs ($W>1$), irrespective of even or odd $W$, reduce to $W=1$ similar to the topological phase in the original Kitaev chain.\\
				 
				 From the previous section we observe different possibilities of superposition of TQCLs. Higher order of neighboring interaction generates higher order WN as well as corresponding higher order TQCLs. The higher order TQCLs are less stable with respect to $\alpha$ and continuously undergo superposition with its lower orders. These higher order TQCLs are comparatively less expressive in their nature and gets dominated by their lower order TQCLs when they undergo superposition. In the meantime some of the higher order TQCL vanish either by creating a gap opening in the energy spectrum or by becoming complex. This results in the suppression of higher order WNs and the system reduces continuously to its lower order and finally to short-range Kitaev chain.\\
				 The inset of of Fig.~\ref{avsm} shows the second derivative of the ground state energy 
				 	\begin{equation}
				 E^{''}(\alpha)=-\frac{1}{2\pi}\int_{-\pi}^{\pi}\frac{\partial^2 E}{\partial \alpha^2}dk,
				 	\end{equation}
				 	with $E$ being the ground state energy, as a function of $\alpha$ (For a many-body system $E=-\sum_{k}E_k$\cite{niu2012majorana,malard2020multicriticality} and the summation is replaced by the integration when it comes to the Brillouin zone limit).\\
 The discontinuity in the 
	derivative of the ground state energy symbolizes the order of the quantum phase transition~\cite{chen2008intrinsic,kempkes2016universalities,cats2018staircase}.  For a topological system, these discontinuities represent the non-analyticities of
	the ground-state energy\cite{niu2012majorana,malard2020multicriticality}.
	Here we observe discontinuity in second 
	order derivative of energy with respect to $\alpha$, which indicates the transitions are second order TQPTs. The maximum among the peaks signifies the TQPT from highest WN to 
	consecutive even/odd WNs depending on even/odd $r$. For the higher order WN, 
	the maximum peak shifts towards the lower  values of $\alpha$ implying the shorter existence of these phases as $\alpha$ increases. We also notice that the amplitudes of peaks shift to higher values of $E^{\prime\prime}$ with increasing number of the interacting neighbors $r$.
\subsection{\label{long}Analysis of long-range model through momentum space characterization}	
\begin{figure}[H]
	\centering
	\includegraphics[width=\columnwidth,height=8cm]{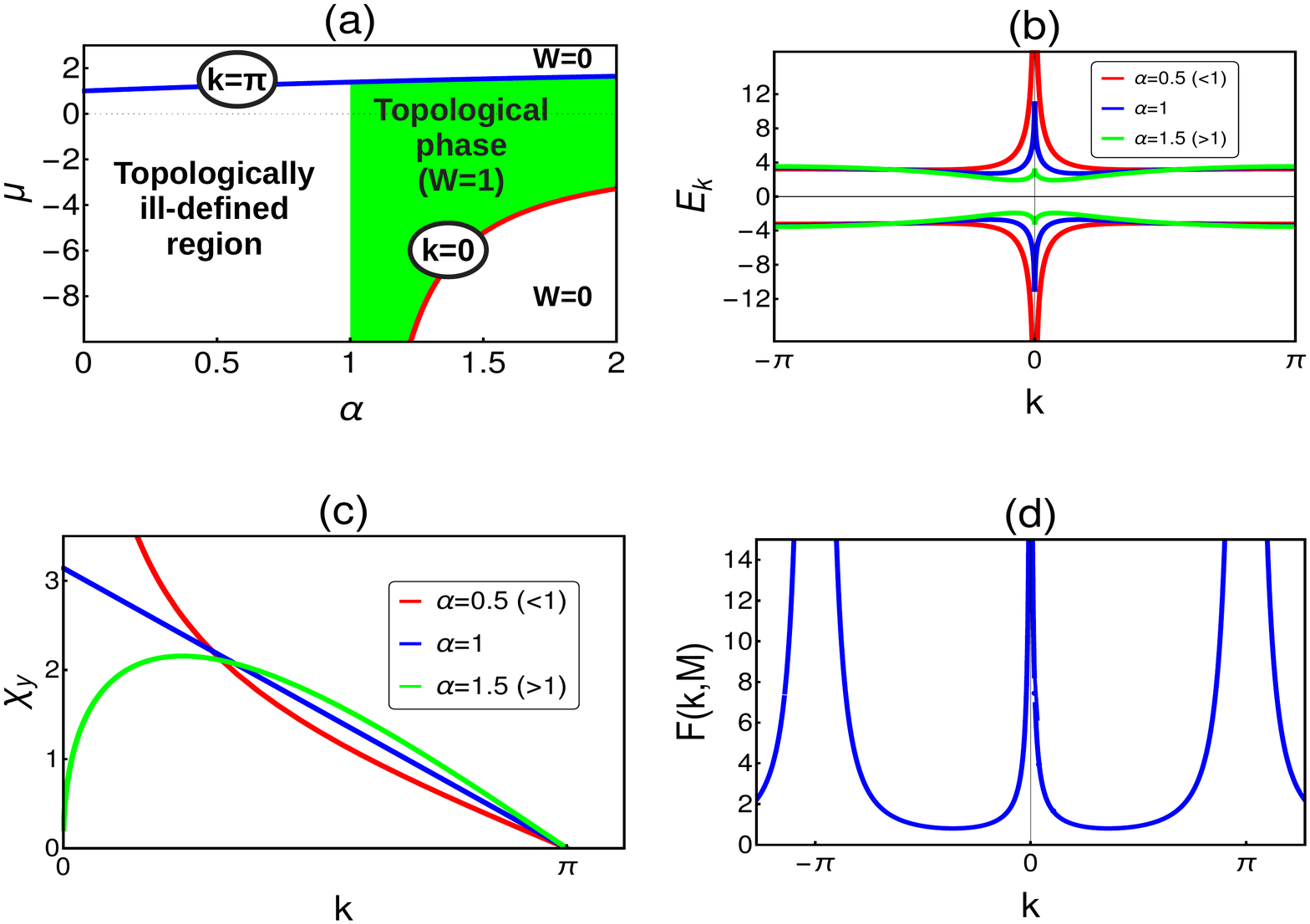}
	\caption{ (a) General phase diagram of isotropic long-range Kitaev chain with $\lambda=1$. (b) Energy dispersion for different values of $\alpha$. The system remains gapped for all values of $\alpha$ and $E_k$ diverges for $\alpha<1$. Here $\mu=-2,\lambda=1$ and system size $l=2000$. (c) Dispersion of the $\chi_y$ parameter at $k=0$ and $k=\pi$ for different values of $\alpha$ with $\mu=0,\lambda=1$ and system size $l=2000$. The term $\chi_y$ diverges, $k\rightarrow0$ as $(k)^{\alpha-1}$ for all $\alpha<1$ and converges for $\alpha>1$. (d) Berry connection plot for $\alpha<1$, which shows the non-analytic behavior through out the $\alpha<1$ region where $\mu=0$ and $\lambda=1$.}
 \label{lr}
\end{figure}

 In previous sections, we have studied the criticality and momentum space characterization of longer-range Kitaev chain with different number of neighbors. In this section we consider Kitaev chain with infinite number of neighbors (i.e. $r\rightarrow\infty$) where the hopping term and the pairing terms decay with parameter $\alpha$ (Eq.~\ref{pseudospin}). The corresponding energy dispersion is given by Eq~\ref{endisp}.
Here both sine and cosine functions lead to polylogarithmic functions, given by
\begin{eqnarray}
\chi_y(k)&=&2\lambda\left( \frac{Li_{\alpha}[e^{i k}]-Li_{\alpha}[e^{-i k}]}{2i}\right) ,\nonumber\\
\chi_z(k)&=&-\mu-2\lambda\left( \frac{Li_{\alpha}[e^{i k}]+Li_{\alpha}[e^{-i k}]}{2}\right) ,\label{pol}
\end{eqnarray}
where we consider an isotopic long-range Kitaev chain for which $J_0=\Delta_0=\lambda$ and $\alpha=\beta$.
The term $\chi_y$ (Eq.~\ref{pol}) decides the number of gap closing points in the BZ and $\chi_z$ decides the parameters to be tuned for the critical conditions i.e., $\mathbf{M}\rightarrow \mathbf{M}_c$.  The term $\Delta_0=\lambda$ in the $\chi_y$ represents the finite superconducting gap and does not influence the gap closing conditions of the system. The parameters $\mu$ and $J_0=\lambda$ act as the tuning parameters ($\mathbf{M}$) and $\chi_z$ gives the critical line which is nothing but the topological phase boundary. These critical lines assure the gap closing conditions during the topological phase transitions. When $\mathbf{M}$ is away from $\mathbf{M}_c$, the system is gapped and with $\mathbf{M}\rightarrow \mathbf{M}_c$ the system attains gapless condition.\\ 
In Eq.~\ref{pol}, pseudo-spin parameters exhibit polylogarithmic nature, where the gap closing ($\chi_y\rightarrow0$) occurs at $k=0$ and $k=\pi$ for different regime of decay parameter as shown in Table~\ref{uc} and corresponding phase diagram is given by Fig.~\ref{lr} (a).
\begin{table}[H]
 \begin{center}
\begin{tabular}{ |c|c|c| } 
\hline
\hline
Decay parameter &$k=0$&$k=\pi$\\
\hline
\hline
When $\alpha>1$&$\mu=-2J_0 \zeta[\alpha]$&$\mu=-2J_0(2^{1-\alpha}-1)\zeta[\alpha]$\\
\hline
When $\alpha<1$&-&$\mu=-2J_0(2^{1-\alpha}-1)\zeta[\alpha]$\\
\hline
\hline
\end{tabular}
\end{center}
\caption{Possible TQCLs for different regimes of an isotropic long-range Kitaev chain. When $\alpha<1$, the TQCL does not exists as $k\rightarrow0$.}
\label{uc}
\end{table}
\noindent When $k=\pi$, criticality occurs ($\mathbf{M}\rightarrow\mathbf{M_c}$) i.e., $E_k\rightarrow0$ at $\mu=-2J_0(2^{1-\alpha}-1)\zeta[\alpha]$  as $\chi_y\rightarrow0$ for all values of $\alpha$.  When $k=0$, criticality occurs at $\mu=-2J_0Li_{\alpha}[1]$, where $\chi_y\rightarrow0$ only for $\alpha>1$.
Hence both $\alpha\leq1$ and $\alpha>1$ belong to two different topological phases without a boundary. Fig.~\ref{lr}(b) shows the energy spectrum of long-range model, where $\alpha<1$ and $\alpha>1$ belonging to two different gapped phases with distinct topological properties, where the transition among them occurs at $\alpha=1$ without gap closing. In a similar model, transition without gap closing has been reported in Ref.~\cite{vodola2014kitaev}. This kind of bifurcation occurs due to the behavior of $\chi_y$ with respect to $\alpha$. $\chi_y$ forms gapless condition at $k=\pi$ irrespective of $\alpha$ but it diverges and fails to make gapless condition at $k=0$ when $\alpha\leq1$ (Fig~\ref{lr} c). However, due to this nature of $\chi_y$, Eq.~\ref{cf1} ($d\theta_k/dk=d(tan^{-1}(\chi_y/\chi_z))/dk$) also becomes non-analytic for $\alpha\leq1$ region even for gapped phases (Fig~\ref{lr} d). Hence it is not possible to define integer WN for $\alpha\leq1$, where $W$ takes positive and negative fractional values below and above the critical line $\mu=-2J_0(2^{1-\alpha}-1)\zeta[\alpha]$ respectively.\\
 This situation can be analyzed by expanding the polylogarithmic function around gap closing points. Expansions of polylogarithmic functions is given by \cite{olver2010nist,vodola2014kitaev},
\begin{equation}
Li_{\alpha}[e^{ik}]=\Gamma[1-\alpha](-ik)^{\alpha-1}+\sum_{n=0}^{\infty}\frac{\zeta[\alpha-n]}{n!}(ik)^n,\label{expansion}
\end{equation}
where $\alpha\neq1,2,3....,|ln(e^{ik})|<2\pi$. Substituting above equation in Eq.~\ref{pol} and after few steps of simplification, we get
\begin{eqnarray}
\chi_y&=&2\lambda(\Gamma[1-\alpha](k)^{\alpha-1}\cos(\frac{\pi\alpha}{2})\nonumber\\
&+&\sum_{n=0}^{\infty}\frac{\zeta[\alpha-n]}{n!}(k)^n\sin(\frac{\pi n}{2})),
\label{poly}
\end{eqnarray}
Thus the $\chi_y$ series diverges if $k\rightarrow0$ as $(k)^{\alpha-1}$ for all $\alpha<1$ and convergence for $\alpha>1$ (Fig~\ref{lr} c). When $\alpha=1$, the polylogarithmic series expansion in Eq.~\ref{poly} is ill-defined as per Eq.~\ref{expansion} \cite{olver2010nist}. 
The physics becomes even more rich in case of Kitaev chain with long-range pairing and short-range hopping \cite{vodola2014kitaev,alecce2017extended}. Because of the conditions imposed, one can get new quasi-particles like massive edge (when $\mathbf{M} \neq \mathbf{M}_c$) modes which are neither zero modes nor can be absorbed by bulk modes. The associated Majorana zero modes are created by $\chi_y$ term ($\mathbf{M}\rightarrow \mathbf{M}_c$) and the massive edge modes are controlled by $\chi_z$ term (which is short-range) \cite{alecce2017extended}. 
These massive edge modes are not prominent in our isotropic Kitaev chain (long-range hopping and long-range pairing) as compared to the case of long-range pairing and short-range hopping. 
However there are similar works which predict the signature of massive edge modes along with Majorana zero modes for $\alpha<1$ region with different approach \cite{alecce2017extended}.\\
\indent As per the known techniques of momentum space characterization, it is not possible to define the topological invariant in gapless regions as well as $\alpha\leq1$ region of a long-range model because of the non-analyticity of the Berry connection.
In the limit $\alpha>1$, the long-range model reduces to the short range version with a Majorana zero mode at each end while in the limit $\alpha\leq1$, there exists an ill-defined region where the BC acquires non-analytic form (Fig.~\ref{lr} d).
Thus the critical exponents of characteristic length as well as susceptibility factors become undefined in this regime.	
\section{A representation of criticality through parameter space}\label{para}
	Here we consider different parameter space and try to analyze the topological phases as well as TQPTs through pseudo-spins as well as from exact calculations.
	\subsection{Pseudo-spin vector parameter space}
\begin{figure}[H]
 \centering
 \includegraphics[width=7cm,height=11cm]{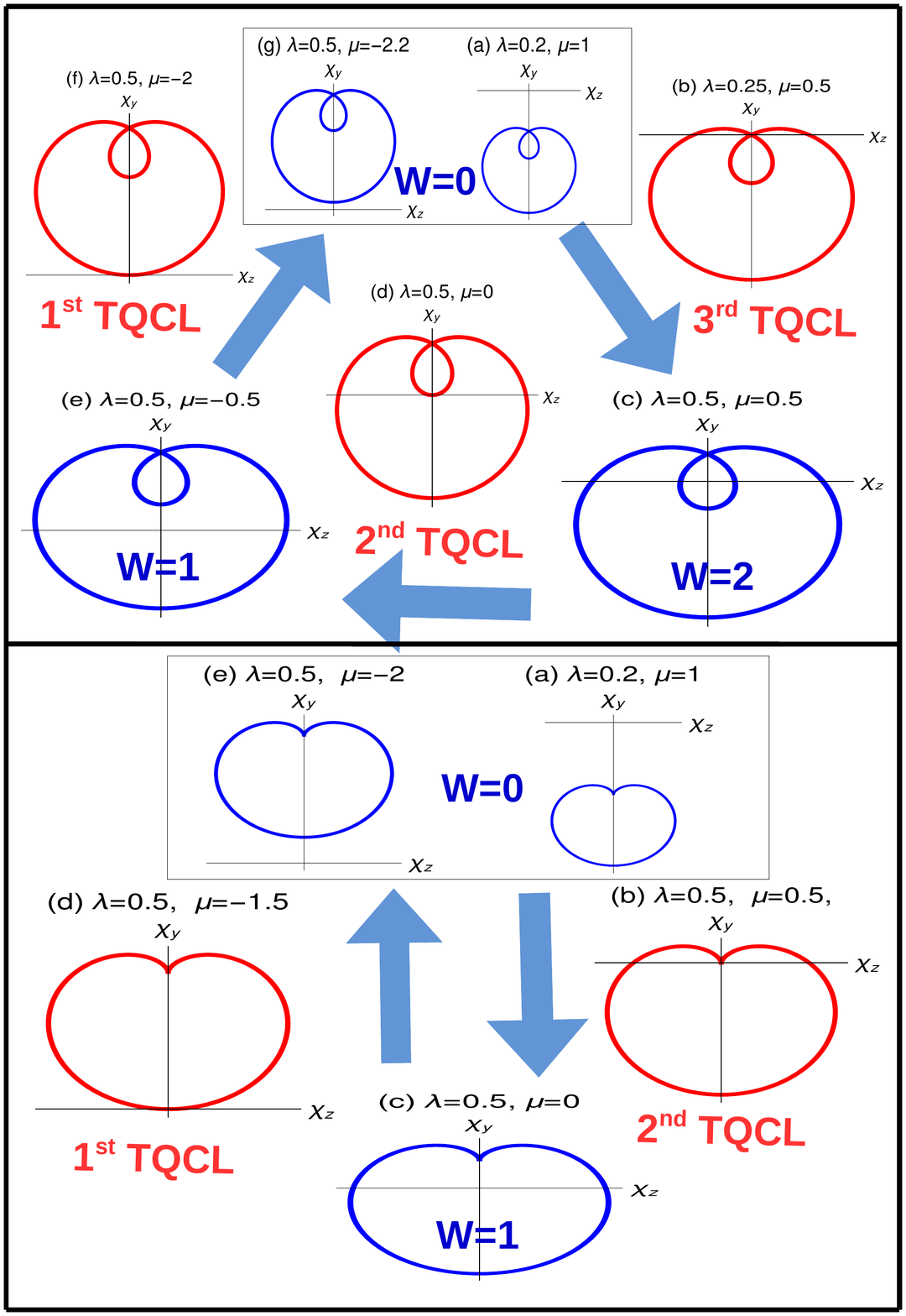}
 \caption{Closed curves representing the pseudo-spin vectors for different values of parameters and for $r=2$. Blue and red curves are topological phases and TQPT respectively. {\itshape Upper panel ($\alpha=0$):} The longer-range Kitaev chain exhibits TQPT among $W:0\leftrightarrow2$ through 3rd TQCL, $W:2\leftrightarrow1$ through 2nd TQCL and $W:1\leftrightarrow0$ through 1st TQCL. {\itshape Lower panel ($\alpha=1$):} The reduced longer-range Kitaev chain exhibits only two TQPTs. From $W:0\leftrightarrow1$ through 2nd TQCL and $W:1\leftrightarrow0$ through 1st TQCL.}
 \label{pc1}
 \end{figure}

\begin{figure}[H]
 	\centering
 	\includegraphics[width=\columnwidth,height=11cm]{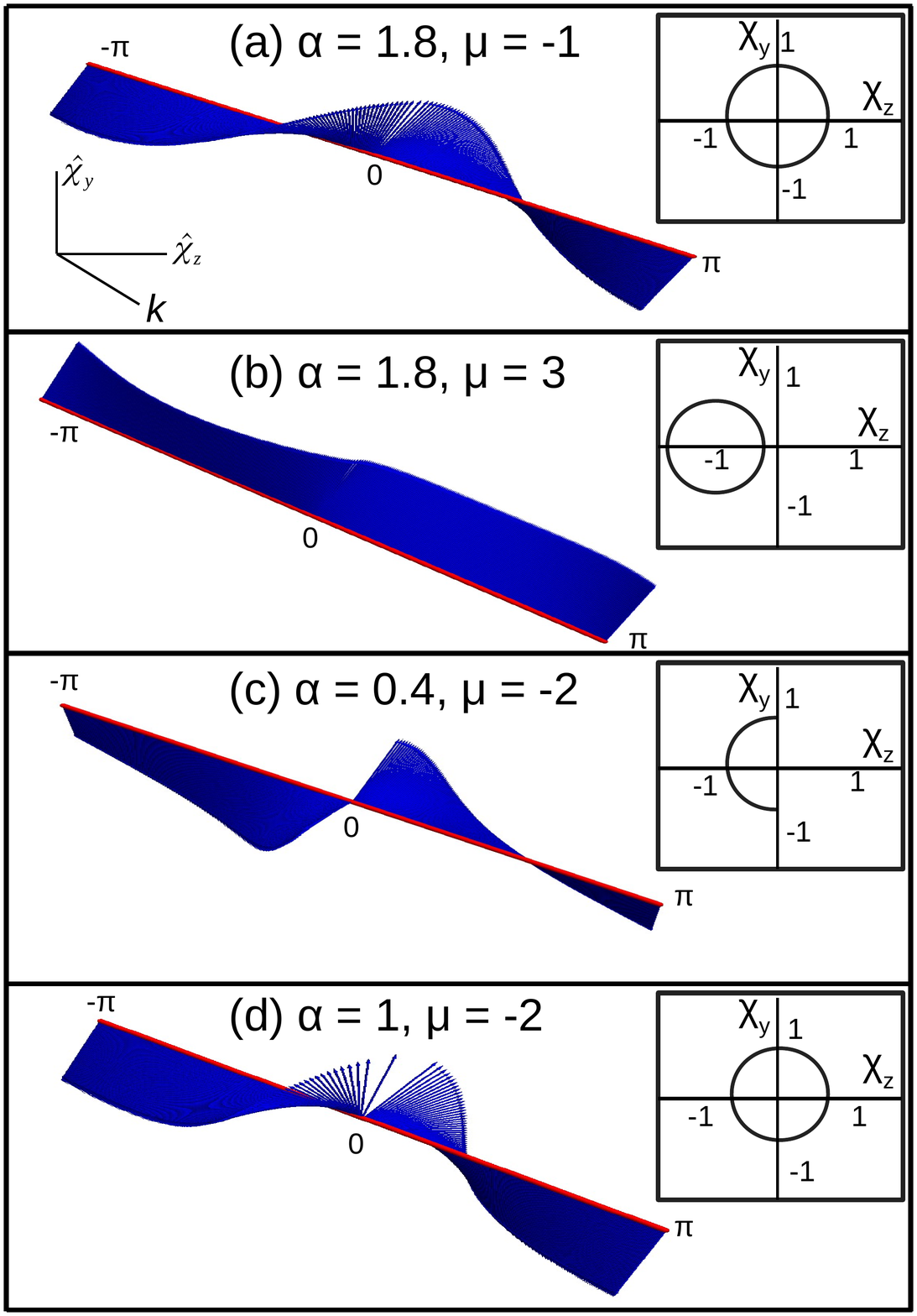}
 	\caption{ Pseudo-spin parameter (unit vector) plots  for different regimes of $\alpha,\mu$ with $\lambda=1$ and system size $l=200$. (a) For $\alpha>1$, the parameter curve completes a full circle around the $k$ axis. (b) For $\alpha>1$, the curve does not completes a circle around the $k$ axis. (c) For $\alpha<1$, the curve shows a discontinuity around $k=0$. (d) For $\alpha=1$, the curve completes a full circle, but contains less populated winding vectors around $k=0$. \textit{Inset:} Corresponding two dimensional representation of parameter plots.}
  \label{ps}
 \end{figure}

Another way to understand the topological properties of 
the system is through the analysis of parameter space~\cite{zhang2015topological,sarkar2018quantization}. The pseudo-spin vector $\vec{\chi} (\chi_x,\chi_y,\chi_z)$ (see Eq.~\ref{pseudospin}) can be used to study the parameter space. The pseudo-spin vector components 
form a closed loop in the 
parameter space due to the periodic boundary condition. If the closed loop encloses the origin, it represents a topological state.
The number of  around the origin gives the topological 
index $W$. If the closed curve does not include the origin, it 
represents non-topological state. When the curve touches 
the origin, it is the critical case where TQPT occurs.\\
 For $r=2$, the components of pseudo-spin vector are given by,
\begin{eqnarray}
\chi_x&=&0,\\
\chi_y&=& 2\lambda(\sin 
(k)+\frac{\sin (2k)}{2^{\alpha}}),\nonumber\\
\chi_z&=&-\mu-2\lambda(\cos (k)+\frac{\cos 
(2k)}{2^{\alpha}}).\nonumber
 \end{eqnarray}
  Fig.~\ref{pc1} represents the pseudo-spin vector in the parameter space for $r=2$. We observe a transition between 
 $W:0\leftrightarrow1$ for different values of $\alpha$.
 The upper panel represents the case when $\alpha=0$. 
 Here we observe transition among $W:0\leftrightarrow2$ through 
 a TQPT (Fig.~\ref{pc1}(b)) followed by a transition between $W:2\leftrightarrow1$ through 
 another TQPT (Fig.~\ref{pc1}(d)). Also we observe a direct transition between $W:1\leftrightarrow0$ trough a third TQPT. We note that for all the TQPTs the closed 
 	curve touches the origin. The lower 
 panel represents $\alpha=1$ case. Here we observe a 
 direct transition between $W:0\leftrightarrow1$ only as we do not find any
 $W=2$ phase.
 When the pseudo-spin curves touch the origin, it represent TQPTs. In general, BC becomes non-analytic at these points and WN becomes ill-defined. However, there are efforts which show the localized edge modes even 
 at the gapless region~\cite{jones2019asymptotic,thorngren2020intrinsically,kestner2011prediction,cheng2011majorana,fidkowski2012majorana,sau2011number,kraus2013majorana,scaffidi2017gapless,jiang2018symmetry}  especially in longer-range models\cite{rahul2019anomalous,kumar2020multi,verresen2019gapless,verresen2018topology,verresen2020topology}. 
 Here we use Eq.~\ref{fw} to find the topological invariant around criticality by taking the limit $k\rightarrow k_0$ as shown in Table~\ref{es}.  						
 We can also observe similar nature in $r=3$ and $r=4$ cases. This is because with the increasing values of $\alpha$, a longer-range model reduces to original Kitaev chain which we have observed also from previous sections.\\
\indent This analysis can also be generalized to a long-range model with some modifications. For a one-dimensional system with periodic boundary conditions, the polylogarithmic functions take the form
 \begin{equation}
Li_{\alpha}[z]=\sum_{l=1}^{\infty}\frac{z^l}{l^{\alpha}}.\label{complex}
 \end{equation}
 which is the polylogarithm of complex function $z$ with order $\alpha$ and for our model $z=e^{ik}\Longrightarrow |z|=1$. Substituting it in Eq.~\ref{pol}, one can get the pseudo-spin parameters in terms of complex function as shown in Fig.~\ref{ps}.
 It is clear from the analysis of section \ref{long}, that the 
 $\alpha>1$ is the topological regime, where the pseudo-spin vector takes a complete rotation around the origin i.e., $W=1$ (Fig.~\ref{ps} a). For the non-topological phase i.e., $W=0$, the pseudo-spin vector fails to take a rotation around the origin  (Fig.~\ref{ps} b). The region $\alpha<1$ is ill-defined from the topological perspective, where we find non-closing curves (Fig.~\ref{ps} c). The point $\alpha=1$ represents a phase transition without gap closing, hence we find a closed curve which does not touches the origin (Fig.~\ref{ps} d). One can obtain the fractional and integer winding numbers for $\alpha<1$ and $\alpha>1$ regime respectively. The point $\alpha=1$ is a case where the pseudo-spin vectors are neither convergent nor divergent (Fig.~\ref{lr} c). Thus we find a closed curve which contains less populated vectors around $k=0$, where the curvature function still behaves as non-analytic resulting in the fractional WN. We observe same behavior of closed curve and less population of vectors around $k=0$ till $\alpha=1.5$. But this regime contains a non-divergent curvature function along with integer winding number. The authors of Ref.~\onlinecite{viyuela2016topological} have used a similar approach of pseudo-spin vector curves to characterize the topological phases in a similar model. Also fractional winding number has been obtained in Ref.~\onlinecite{alecce2017extended} in the same regime for a very similar model.
\subsection{A few exact solutions for topological characterization}
 Here we try to find some exact solutions of the winding 
  number for different parameters. It is 
  difficult to find exact solutions for the parameter 
  spaces as the associated integral may become really complicated. We consider only some of the special cases and 
  present the exact solutions of WN for $r=2$ only.
  Eq.~\ref{wn1} gives the expression of 
     WN for $r=2$. 
     WN always gives integer number for gapped topological phases. But for gapless phases WN may take integer or fraction values depending on the parameter space. For a gapless phase we calculate WN by omitting the gap closing points
      i.e., $k=0,\pi$ and $\cos^{-1}(-2^{\alpha-1})$. Here the WN takes the form,
      \begin{equation}
      W=\int\frac{A+B\cos(k)+C\cos(2k)}{a+b\cos(k)+c\cos(2k)},
       \end{equation}\label{ex}
      where 
     \begin{eqnarray}
     A&=&4\lambda^2(2+2^{2\alpha}),B=2^{\alpha+1}\lambda(6\lambda+2^{\alpha}\mu),\nonumber\\
     C&=&c=2^{\alpha+2}\lambda\mu,a=4\lambda^2(1+2^{\alpha})2^{2\alpha}\mu^2,\nonumber\\
     b&=&2^{\alpha+2}\lambda(2^{\alpha}\mu+2\lambda^2)\nonumber
     \end{eqnarray}
 Based on parameter space, WN takes different standard integral formats \cite{sarkar2018quantization,olver2010nist,chen2019topological}. 
         
\begin{widetext}

\begin{table}[H]
\begin{center}

\begin{tabular}{ |c|c|c|c|c| } 
\hline
Phase&$\hspace{0.2cm}\alpha\hspace{0.2cm}$&Relation& Expression& Winding Number \\ 
\hline
\hline
&&&&\\
1st TQCL ($k=0$) &0&  $\mu=-4\lambda$&$W=\left( \frac{1}{2\pi}\right) \int_{-\pi}^{\pi}\frac{7+8\cos(k)}{10+8\cos(k)}dk$ & W=1/2\\
($\mu=-2\lambda(1+1/2^{\alpha})$)&&&&\\
 &1&  $\mu=-3\lambda$&$W=\left( \frac{1}{2\pi}\right) \int_{-\pi}^{\pi}1-\frac{2}{5+3\cos(k)}dk$ & W=1/2\\
\hline
&&&&\\		 
2nd TQCL ($k=\pi$) &0&  $\mu=0$&$W=\left( \frac{1}{2\pi}\right) \int_{-\pi}^{\pi}3/2dk$ & W=3/2\\
($\mu=-2\lambda(-1+1/2^{\alpha})$)&&&&\\
 &1&  $\mu=\lambda$&$W=\left( \frac{1}{2\pi}\right) \int_{-\pi}^{\pi}dk$ & W=1\\
\hline
&&&&\\
3rd TQCL ($k=\cos^{-1}(-2^{\alpha-1})$) &0&  $\mu=2\lambda$&$W=\left( \frac{1}{2\pi}\right) \int_{-\pi}^{\pi}dk$ & W=1\\
($\mu=\lambda/2^{\alpha-1}$)&&&&\\		
 &1&  $\mu=\lambda$&$W=\left( \frac{1}{2\pi}\right) \int_{-\pi}^{\pi}dk$ & W=1\\		 		 
\hline
\hline
&&&&\\
Gapped Phase & 0&  $\mu=-\lambda$&$W=lim_{\delta\longrightarrow0^+}\left( \frac{1}{2\pi}\right) \int_{\pi-\delta}^{\pi+\delta}\frac{3+6\cos(k)}{9+4\cos(k)-4\cos(2k)}dk$ & W=1\\
\hline
&&&&\\
Gapped Phase &1&  $\mu=-\lambda$&$W=lim_{\delta\longrightarrow0^+}\left( \frac{1}{2\pi}\right)\int_{\pi-\delta}^{\pi+\delta}1-\frac{2\cos(k)}{-3+\cos(2k)}dk$ & W=1\\
\hline
&&&&\\
Gapped Phase &0&  $\mu=\lambda$&$W=lim_{\delta\longrightarrow0^+}\left( \frac{1}{2\pi}\right)\int_{\pi-\delta}^{\pi+\delta}\frac{3+2\cos(k)}{9+12\cos(k)+4\cos(2k)}dk$ & W=2\\
\hline
&&&&\\
Gapped Phase & 1&  $\mu=\lambda$&$W=lim_{\delta\longrightarrow0^+}\left( \frac{1}{2\pi}\right)\int_{\pi-\delta}^{\pi+\delta}dk$ & W=1\\

\hline
\hline
\end{tabular}

\end{center}
\caption{A few exact solutions for the winding number when $r=2$.}
	\label{es}
\end{table}

\end{widetext}
Some of the standard integral formats are,\\
          1) $\int\frac{A+B \cos k}{a+b \cos k}dk$ when $a^2>b^2,(C=c=0)$
          \begin{equation}
         =\frac{B}{b}k+\frac{2(Ab-aB)}{b\sqrt{a^2-b^2}}\tan^{-1}(\frac{\sqrt{a^2-b^2}\tan k/2}{a+b})
          \end{equation}
          2)  $\int\frac{A+B \cos k}{a+b \cos k}dk$ when $b^2>a^2,(C=c=0)$
          \begin{equation}
          =\frac{B}{b}k+\frac{2(Ab-aB)}{b\sqrt{b^2-a^2}}Log(\frac{\sqrt{b^2-a^2}\tan k/2+(a+b)}{\sqrt{b^2-a^2}\tan k/2-(a+b)})
           \end{equation}
         Other than this we have couple of more cases,\\\\
         3) Consider $z=\exp(ik)$ and $dk=dz/iz$,
         \begin{equation}
         \int\frac{A dk}{a+b\cos(k)}=A\oint\frac{dz/iz}{a+b(z+1/z)/2}=8\pi A(z_+-z_-)/b,
         \end{equation}
         where $z_{\pm}=-a/b\pm\sqrt{a^2/b^2-1}$. Here $z_+(z_-)$ is defined as the root that is inside (outside) the contour $|z|=1$.\\\\
         4) In the similar way,
         \begin{equation}
         \int\frac{B\cos(k)dk}{a+b\cos(k)}=\left(\frac{z_+}{z_+-z_-}+\frac{1}{z_+z_-}+\frac{1}{z_+(z_+-z_-)}\right).
         \end{equation}
         Based on parameter space, WN gets different quantized 
         values. Here we consider a few cases (see Table~\ref{es}). \\
 \indent For a topological system WN represents the number localized edge modes of the gapped phases. Based on the number of interacting neighbors $r$, here we get corresponding exact solutions for the topological phases. But at the criticality the exact solutions are calculated by omitting gap closing points. For the case $r=2$, 1st and 3rd TQCLs give fractional as well as integer exact solutions respectively for all parameter space. The 2nd TQCL gives fractional solutions for initial values and integer solution as $\alpha\rightarrow\infty$ (Table~\ref{es}).
\section{Outlook and experimental possibilities}\label{gen}
 In this work we have used the isotropic conditions 
($\alpha=\beta$, $J_0=\Delta_0=\lambda$) and longer-range as well as long-range model  
to explain topological characterization and criticality of the model. The results of our 
work can be generalized to other parameter space also. When 
$\alpha,\beta\rightarrow\infty$, the model reduces to 
original Kitaev model. But through our work we realize that, the reduction should undergo through the
process of superposition and vanishing of TQCL as discussed earlier. 
According to the available literature, it is clear that winding number is not enough to understand the topological properties of a long-range model. And there are two different arguments about the reduction of long-range model to short-range model. Some works suggest that the reduction happens when $\alpha>1$~ \cite{lepori2018edge,giuliano2018current,lepori2017long} and some studies obtain short-range limit for $\alpha >1.5$ ~\cite{viyuela2016topological}. Solving these issues through the study of universality class of critical exponents and CFT can give the better understanding.
When 
the pairing term decays slower than hopping parameter, 
there may be possibilities of obtaining new exotic 
particles like massive Majorana modes~\cite{vodola2015long}. The results of criticality and behavior of TQCLs may be 
interesting in those cases.\\
\noindent{\it Experimental Possibilities:} There are a number 
of experiments which explore the properties of 
long-range models especially in trapped ions~\cite{deng2005effective,britton2012engineered,hauke2010complete,roy2019quantum},
 atom coupled to multi-mode cavities~\cite{douglas2015quantum}, magnetic impurities~\cite{zhang2019majorana,menard2015long} and quantum 
computation~\cite{amin2019information}. In long-range 
models, the characteristic length shrinks for the longer 
neighbors. Hence, even by using a relatively small 
number of ions it is possible to suppress the 
finite-size effects ~\cite{gong2016topological}. Within the tight-binding BdG 
formalism, the Shiba chains can be modeled to p-wave 
superconducting Kitaev chain with long-range pairing and 
hopping~\cite{rontynen2015topological,pientka2014unconventional}.
 Naturally Shiba chains exhibits $1/r$ decay away from 
 certain limits of coherence length~\cite{gong2016topological}. Hence it is easy to map our 
 isotropic Kitaev model in such systems. Analysis of criticality studied in this work 
 may help to explore the subject in a better 
 way. \\
\section{Conclusion}\label{conclusion}
 To summarize, we have presented a theoretical study of the topological quantum phase transitions and quantum criticality in the longer-range as well as long-range Kitaev chain. 
Here all possible topological criticality conditions have been calculated in detail along with precise topological phase diagram. For a longer-range model, with the increasing number of interacting neighbors, higher order winding numbers have been generated and their stability decreases with the increasing value of the decay parameter, which has been verified by the analysis of ground-state energy. A decrease of winding number has been observed with decreasing long-rangeness in the system with a pattern of odd-to-odd and even-to-even transition among winding numbers. As a reason behind this, we show a 
mechanism of the superposition of two critical lines and vanishing of the one with higher winding number. We have analyzed different possibilities of superposition and different means of vanishing of critical lines. Through this we have studied the nature of multi-critical points in the longer-range model. The criticality has been studied from the perspective of critical exponents and their fate near the multi-critical points have been analyzed. As the generalization of this work, we have considered a truly long-range model and its momentum space characterization been done. We have analyzed the non-analytic behavior of Berry connection in both long-range Kitaev chain and longer-range (finite-range) Kitaev chain to study the topological invariant.
A parameter space representation is done for longer-range as well as long-range models along with a few exact solutions for the winding numbers in support of our findings. 
We have also discussed the possible outlook and experimental aspects of our work.
Instances of study of topological quantum phase transitions covering long-range models and quantum criticality are rare in the literature. We hope that our work will help boost the understanding of such systems.\\\\
\textbf{Acknowledgments:}
    SS would like to acknowledge DST (EMR/2017/000898) 
    for the funding and RRI and ICTS library for the books and 
    journals. NR is grateful to the University Grants 
    Commission (UGC), India for providing a PhD 
    fellowship. YRK would like to thank Admar Mutt Education Foundation for the scholarship. The authors would like to acknowledge Dr. B S Ramachandra,
    Prof. Prabir K. Mukherjee and Prof. C Sivaram who read 
    this manuscript critically and gave useful 
    suggestions.
    \bibliography{LongRange}
  
  \appendix

\section{Detailed derivation of critical lines for longer-range Kitaev chain}\label{appC}
Here we derive the topological quantum critical lines of the longer-range Kitaev chain for interacting neighbors $r=2,3$ and $4$. We follow the standard method to find the critical lines. First we find the value of $k_0$ by making the term $\chi_y=0$. After that we substitute the value of $k_0$ into $\chi_z=0$ to find the critical line. Depending on the number of interacting neighbors, that many critical lines will be generated.
\subsubsection*{When r=2}
\noindent Here the gap closing occurs for three different values of $k$, i.e.\\
\begin{eqnarray}
1) \hspace{0.5cm}k&=&0,\nonumber\\
2) \hspace{0.5cm}k&=&\pi,\nonumber \\
3) \hspace{0.5cm}k&=&\cos^{-1}(-2^{\alpha-1})\nonumber.
\end{eqnarray}
 Corresponding TQCLs are,
\begin{eqnarray}
1) \hspace{0.5cm}\mu&=&-2\lambda(\frac{1}{2^{\alpha}}+1),\nonumber\\
2) \hspace{0.5cm}\mu&=&-2\lambda(\frac{1}{2^{\alpha}}-1),\nonumber\\
3)\hspace{0.5cm} \mu&=&\frac{\lambda}{2^{\alpha-1}}.\nonumber
\end{eqnarray}
for 3rd TQCL $0<\alpha\leq1.$
\subsubsection*{When r=3}
\noindent Here the gap closing occurs for four different values of $k$, i.e.\\
\begin{eqnarray}
1)\hspace{0.5cm} k&=&0,\nonumber\\\nonumber\\
2)\hspace{0.5cm} k&=&\pi,\nonumber\\\nonumber\\
3)\hspace{0.5cm} k&=&\cos^{-1}(\frac{3^{\alpha}}{4}(-\frac{1}{2}+\sqrt{\frac{1}{2^{\alpha}}-\frac{4(3^{\alpha}-1)}{3^{2\alpha}}})),\nonumber\\\nonumber\\
4)\hspace{0.5cm} k&=&\cos^{-1}(\frac{3^{\alpha}}{4}(-\frac{1}{2}-\sqrt{\frac{1}{2^{\alpha}}-\frac{4(3^{\alpha}-1)}{3^{2\alpha}}})).\nonumber
\end{eqnarray}
\text{ Corresponding TQCLs are,}
\begin{eqnarray}
1)\hspace{0.5cm} \mu&=&-2\lambda\left(\frac{1}{3^{\alpha }}+\frac{1}{2^{\alpha }}+1\right),\nonumber\\\nonumber\\
2) \hspace{0.5cm}\mu&=&-2 \lambda\left(-\frac{1}{3^{\alpha }}+\frac{1}{2^{\alpha }}-1\right),\nonumber\\\nonumber\\
3)\hspace{0.5cm} \mu&=&-\lambda(-8^{-\alpha -1} Q),\nonumber\\\nonumber\\
4) \hspace{0.5cm}\mu&=&-\lambda(8^{-\alpha -1} R).\nonumber
\end{eqnarray}
where,
\begin{eqnarray}
P&=&\sqrt{2^{-\alpha }-4\times9^{-\alpha } \left(3^{\alpha }-1\right)},\nonumber\\
Q&=&\left(18^{\alpha } P -36^{\alpha } P+8^{\alpha +1} P+9^{\alpha }-18^{\alpha }-2^{2 \alpha +3}\right),\nonumber\\
R&=&\left(18^{\alpha } P -36^{\alpha } P+8^{\alpha +1} P-9^{\alpha }+18^{\alpha }+2^{2 \alpha +3}\right).\nonumber
\end{eqnarray}
\subsubsection*{When r=4}
\noindent Here the gap closing occurs for five different values of $k$, i.e.\\
\begin{eqnarray}
&1)& \hspace{0.5cm}k=0,\nonumber\\\nonumber\\\nonumber\\
&2)& \hspace{0.5cm}k=\pi,\nonumber\\\nonumber\\\nonumber\\
&3)& \hspace{0.5cm}\cos(k)=-2^{2 \alpha -1}\times3^{-\alpha -1}\nonumber\\&+&\frac{1}{3} 2^{2 \alpha -\frac{10}{3}} \sqrt[3]{\sqrt{4 A^3+X^2}+X}\nonumber\\
&-&\frac{2^{2 \alpha -\frac{8}{3}} A}{3 \sqrt[3]{\sqrt{4 A^3+X^2}+X}},\nonumber\\\nonumber\\\nonumber\\
&4)& \hspace{0.5cm}\cos(k)=-2^{2 \alpha -1}\times3^{-\alpha -1}\nonumber\\&-&\frac{1}{3} \left(1-i \sqrt{3}\right) 2^{2 \alpha -\frac{13}{3}} \sqrt[3]{\sqrt{4 A^3+X^2}+X}\nonumber\\
&+&\frac{\left(1+i \sqrt{3}\right) 2^{2 \alpha -\frac{11}{3}} A}{3 \sqrt[3]{\sqrt{4 A^3+X^2}+X}},\nonumber\\\nonumber\\\nonumber\\
&5)& \hspace{0.5cm}\cos(k)= -2^{2 \alpha -1}\times3^{-\alpha -1}\nonumber\\
&-&\frac{1}{3} \left(1+i \sqrt{3}\right) 2^{2 \alpha -\frac{13}{3}} \sqrt[3]{\sqrt{4 A^3+X^2}+X}\nonumber\\
&+&\frac{\left(1-i \sqrt{3}\right) 2^{2 \alpha -\frac{11}{3}} A}{3 \sqrt[3]{\sqrt{4 A^3+X^2}+X}}.\nonumber
\end{eqnarray}
where
\begin{eqnarray}
A&=&3\times2^{4-4 \alpha } \left(2^{\alpha }-2\right)-16\times 3^{-2 \alpha },\nonumber\\
X&=&-27\times 2^{6-4 \alpha }+2^{6-4 \alpha } 3^{2-\alpha }+2^{6-3 \alpha } 3^{2-\alpha }\nonumber\\
&-&128\times3^{-3 \alpha },\nonumber\\
P&=&\sqrt{4 A^3+X^2}+X,\nonumber\\
Q&=&-2^{2 \alpha -1} 3^{-\alpha -1}-\frac{2^{2 \alpha -\frac{8}{3}} A}{3 \sqrt[3]{P}}+\frac{1}{3} 2^{2 \alpha -\frac{10}{3}} \sqrt[3]{P},\nonumber\\
R&=&-2^{2 \alpha -1} 3^{-\alpha -1}+\frac{\left(1-i \sqrt{3}\right) 2^{2 \alpha -\frac{11}{3}} A}{3 \sqrt[3]{P}}\nonumber\\
&-&\frac{1}{3} \left(1+i \sqrt{3}\right) 2^{2 \alpha -\frac{13}{3}} P,\nonumber\\
S&=&-2^{2 \alpha -1} 3^{-\alpha -1}+\frac{\left(1+i \sqrt{3}\right) 2^{2 \alpha -\frac{11}{3}} A}{3 \sqrt[3]{P}}\nonumber\\
&-&\frac{1}{3} \left(1-i \sqrt{3}\right) 2^{2 \alpha -\frac{13}{3}} P.\nonumber
\end{eqnarray}
\\
Corresponding TQCLs are,
\begin{eqnarray}
1)\hspace{0.5cm} \mu&=&-2\lambda \left(\frac{1}{3^{\alpha }}+\frac{1}{4^{\alpha }}+\frac{1}{2^{\alpha }}+1\right),\nonumber\\\nonumber\\
2) \hspace{0.5cm}\mu&=&-2\lambda \left(-\frac{1}{3^{\alpha }}+\frac{1}{4^{\alpha }}+\frac{1}{2^{\alpha }}-1\right),\nonumber\\\nonumber\\
3) \hspace{0.5cm}\mu&=&-2\lambda \left(-2^{1-\alpha }+4^{-\alpha }+2^{3-2 \alpha } Q^4\right)\nonumber\\
&-&2\lambda \left(4\times3^{-\alpha } Q^3
+\left(2^{1-\alpha }-2^{3-2 \alpha }\right) Q^2+\left(1-3^{1-\alpha }\right) Q\right),\nonumber\\\nonumber\\
4)\hspace{0.5cm} \mu&=&-2\lambda   \left(-2^{1-\alpha }+4^{-\alpha }+2^{3-2 \alpha } S^4+4\times3^{-\alpha } S^3\right)\nonumber\\
&-&2\lambda   \left(\left(2^{1-\alpha }-2^{3-2 \alpha }\right) S^2+\left(1-3^{1-\alpha }\right) S\right),\nonumber\\\nonumber\\
5)\hspace{0.5cm} \mu&=&-2\lambda   \left(-2^{1-\alpha }+4^{-\alpha }+2^{3-2 \alpha } R^4+4\times 3^{-\alpha } R^3\right)\nonumber\\
&-&2\lambda   \left(\left(2^{1-\alpha }-2^{3-2 \alpha }\right) R^2+\left(1-3^{1-\alpha }\right) R\right).\nonumber
\end{eqnarray}
\end{document}